\documentclass[12pt]{reportj}
\usepackage{deluxetablej}
\usepackage{hyperref} 
\makeatletter
\def\@to{to}
\makeatother
\usepackage{times}
\usepackage{graphicx}
\usepackage{tocloft}
\usepackage{fancyheadings}
\usepackage{xcolor}

\emergencystretch=\maxdimen
\usepackage{datetime}
\newdateformat{ddmonthyyyy}{\THEDAY\ \monthname[\THEMONTH]\ \THEYEAR}

\renewcommand\thesection{\arabic{section}}
\renewcommand\thesubsection{\thesection.\arabic{subsection}}

\def\ssection#1{\setcounter{subsection}{0} \refstepcounter{section} \section*{\hbox to \hsize{\large\bf \arabic{section}. #1\hfill }}\label{sec} \addcontentsline{toc}{section}{\arabic{section}. #1}}
\def\ssubsection#1{\setcounter{subsubsection}{0} \refstepcounter{subsection}\subsection*{\hbox to \hsize{\normalsize\bfseries\itshape \arabic{section}.\arabic{subsection} #1\hfill}}\label{subsec} \addcontentsline{toc}{subsection}{\arabic{section}.\arabic{subsection} #1}}
\def\ssubsubsection#1{\refstepcounter{subsubsection}\subsection*{\hbox to \hsize{\normalsize\it \arabic{section}.\arabic{subsection}.\arabic{subsubsection} #1\hfill}}\label{subsubsec} \addcontentsline{toc}{subsubsection}{\arabic{section}.\arabic{subsection}.\arabic{subsubsection} #1}}

\def\ssectionstar#1{\section*{\hbox to \hsize{\large\bf #1\hfill}} \addcontentsline{toc}{section}{#1}}
\def\ssubsectionstar#1{\subsection*{\hbox to \hsize{\normalsize\bfseries\itshape #1\hfill}} \addcontentsline{toc}{subsection}{#1}}
\def\ssubsubsectionstar#1{\subsection*{\hbox to \hsize{\normalsize\it  #1\hfill}} \addcontentsline{toc}{subsection}{#1}}

\renewcommand{\cftaftertoctitle}{%
\mbox{}\hfill{\normalfont Page}}
\newcommand{\GB}{G191-B2B}  
\newcommand{\GDseven}{GD~71} 
\newcommand{\GDone}{GD~153} 
\newcommand{\calstis}{CALSTIS}

\begin{document}

~\\

\vspace{-2.4cm}
\noindent\includegraphics*[width=0.295\linewidth]{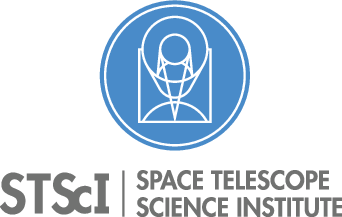}

\vspace{-0.4cm}

\begin{flushright}
    {\bf Instrument Science Report STIS 2025-02(v1)}
    
    \vspace{1.1cm}
    
    {\bf\Huge Updated Sensitivities of the Five STIS L-mode Gratings}
    
    \rule{0.25\linewidth}{0.5pt}
    
    \vspace{0.5cm}
    
    Amy M. Jones$^1$, Svea Hernandez$^1$, Joleen K.\ Carlberg$^1$, Daniel Welty$^1$
    \linebreak
    \newline
    \footnotesize{$^1$ Space Telescope Science Institute, Baltimore, MD\\}
    
    \vspace{0.5cm}
    
     \ddmonthyyyy{30 July 2025}
\end{flushright}

\vspace{0.1cm}

\noindent\rule{\linewidth}{1.0pt}
\noindent{\bf A{\footnotesize BSTRACT}}

\noindent
Re-derivation of the sensitivities of all of the Space Telescope Imaging Spectrograph (STIS) observing modes were required after major updates were introduced to the model atmospheres of the three primary standard stars. The new predicted continuum fluxes were up to 2--3\% different from the models used to originally calibrate STIS.  This work focuses on the re-derivation of spectral sensitivities for the five STIS low-resolution (L-mode) gratings: G140L, G230L, G230LB, G430L, and G750L, which span wavelengths from the far-ultraviolet through the near infrared.  Updated photometric throughput tables were delivered to the Calibration Reference Data System (CRDS) on April 7, 2022 and April 14, 2023, which triggered a recalibration of all historical STIS datasets taken with these modes. The sensitivities derived from each of the standard stars typically agree with one another to better than 1\%, though discrepancies as large as 1.5\% are found in spectral regions most impacted by hydrogen absorption.

\vspace{-0.1cm}
\noindent\rule{\linewidth}{1.0pt}

\renewcommand{\cftaftertoctitle}{\thispagestyle{fancy}}
\tableofcontents


\vspace{-0.3cm}
\ssection{Introduction}\label{sec:Introduction}
The absolute flux calibration of the Space Telescope Imaging Spectrograph (STIS) relies on the observations and theoretical models of three primary standard stars: the white dwarfs \GB, \GDseven, and \GDone\ (\href{https://ui.adsabs.harvard.edu/abs/1995AJ....110.1316B/abstract}{Bohlin, et~al.\ 1995}).   
Calibrated STIS data from the low resolution gratings, in turn, form the observational backbone of the CALSPEC flux standard library (\href{https://ui.adsabs.harvard.edu/abs/2014PASP..126..711B/abstract}{Bohlin et~al.\ 2014}).  As advances are made in the theoretical understanding of these primary stars, all data in the CALSPEC library are updated to a self-consistent, state-of-the-art flux scale. See, e.g., the summary of updates on the main CALSPEC website\footnote{\url{https://www.stsci.edu/hst/instrumentation/reference-data-for-calibration-and-tools/astronomical-catalogs/calspec}}. It is important to note that the STIS data used in the CALSPEC library are reduced with independent software to achieve the highest possible flux precision for those particular data sets and thus can deviate from the standard calibrated data products found in the Barbara A. Mikulski Archive for Space Telescopes (MAST). For a more detailed discussion and comparisons, see \href{https://ui.adsabs.harvard.edu/abs/2019AJ....158..211B/abstract}{Bohlin et~al.\ 2019}. 

STIS data products delivered by MAST are reduced with the \calstis\ pipeline together with reference files managed by the HST Calibration Reference Data System (CRDS). 
Changes to either are relatively slow to be adopted as impacts must be tested on the full archive of legacy STIS data spanning the myriad of available observing modes under a variety of observing conditions (e.g., variations in signal-to-noise (S/N), aperture choices, etc.). Therefore, while the primary standard stars have over half a dozen generations of stellar models available, the release of the CALSPEC version 11 models (hereafter v11, described in Section \ref{sec:mod}) represent the first time the model changes were significant enough to warrant a re-derivation of the sensitivities of STIS's observing modes. Companion reports document updates to modes not covered in this work, including \href{https://ui.adsabs.harvard.edu/abs/2022stis.rept....4C/abstract}{Carlberg et al.\ (2022; E140M)}, \href{https://ui.adsabs.harvard.edu/abs/2024stis.rept....2S/abstract}{Siebert et al.\ (2024; prioritized pre-SM4 echelle modes)}, \href{https://ui.adsabs.harvard.edu/abs/2024stis.rept....4H/abstract}{Hernandez et al. (2024; prioritized post-SM4 echelle modes)}, Fullerton (2025; first-order M modes), and dos Santos et al. (in prep, NUV imaging modes). The status of updated reference file delivery is also tracked on a dedicated website\footnote{\url{https://www.stsci.edu/hst/instrumentation/stis/flux-recalibration}} maintained by the STIS team.

This ISR documents the re-derivation of sensitivities and delivery of \texttt{PHOTTAB} reference files for the five low-resolution (L-mode) STIS gratings.
When used together, the L-mode gratings provide a continuous spectral energy distribution (SED) of an object of interest from 1150~\AA\ -- 1.027~$\mu$m with resolutions ranging from 500 to 1440,  and they are  heavily utilized in the CALSPEC library. 
The previous \texttt{PHOTTAB} reference files used sensitivities derived nearly two decades ago.
The relative changes in these CALSPEC models are summarized in Section \ref{sec:mod}.  The details of the observations and the sensitivity derivations are covered in Sections \ref{sec:obs} -- \ref{sec:G750L}. We test the resultant flux distributions of standard star data reduced with \calstis\  and new \texttt{PHOTTAB} files relative to the model fluxes in Section \ref{sec}, and on average we find $<1$\% residual discrepancies with the models. 


\lhead{}
\rhead{}
\cfoot{\rm {\hspace{-1.9cm} Instrument Science Report STIS 2025-02(v1) Page \thepage}}

\vspace{-0.3cm}
\ssection{CALSPEC Model Updates}\label{sec:mod}

The main motivation for updating the sensitivity of STIS modes is the few percent improvement in the underlying flux models of the three primary white dwarf standard stars, namely \GB, \GDone, and \GDseven. These changes encompass advances in the non-local thermodynamical equilibrium (NLTE) modeling of white dwarf atmospheres, updated stellar effective temperature and surface gravity measurements, and an update to the Vega flux used to normalize the models (\href{https://ui.adsabs.harvard.edu/abs/2020AJ....160...21B/abstract}{Bohlin et al 2020}). Most of the HST instruments are updating or have updated their sensitivities with the newer CALSPEC models, and these models are labeled as \texttt{\_mod011} in the CALSPEC library (note that CALSPEC has released v12 as of the time of writing, but the differences between v11 and v12 are minimal). 
For \GDone\ and \GDseven, the v11 models are pure hydrogen models from the TMAP2019 grid. The \GB\ model is a TLUSTY207 line-blanketed model using \href{https://ui.adsabs.harvard.edu/abs/2013A%26A...560A.106R/abstract}{Rauch et~al.\ (2013)} metal abundances. For the STIS L-modes,  the previous pipeline sensitivities were derived using a combination of v4 (\texttt{\_mod004}, for \GDone\ and \GB) and v5 (\texttt{\_mod005}, for \GDseven) CALSPEC models. These older models relied on TLUSTY pure hydrogen NLTE models for all three primary stars (\href{https://ui.adsabs.harvard.edu/abs/2004AJ....128.3053B/abstract}{Bohlin \& Gilliland 2004}, \href{https://www.stsci.edu/files/live/sites/www/files/home/events/event-assets/2002/_documents/2002-hst-calibration-workshop-presentation-bohlin.pdf}{Bohlin 2002}).

\begin{figure}[!ht]
  \centering
  \includegraphics[scale=0.9]{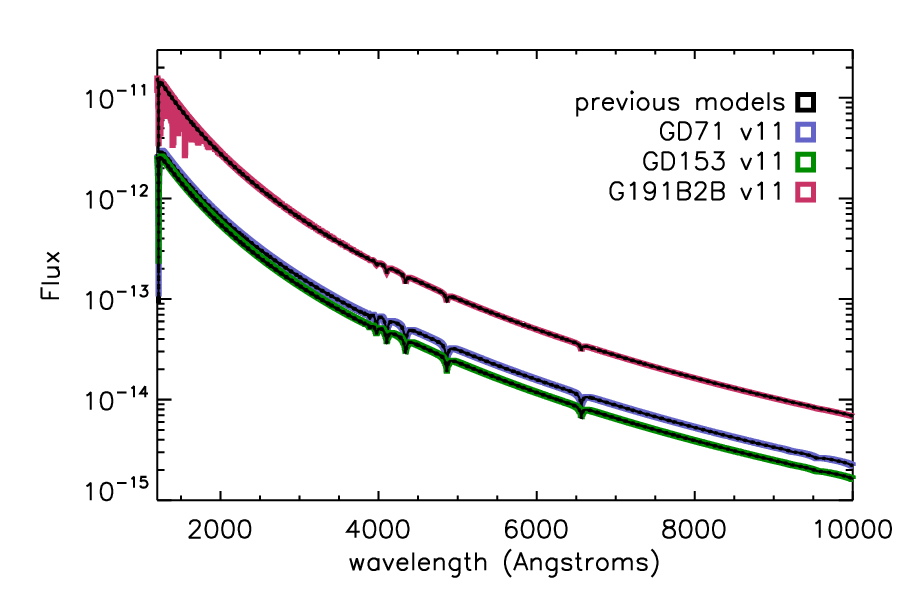}
    \includegraphics[scale=0.9]{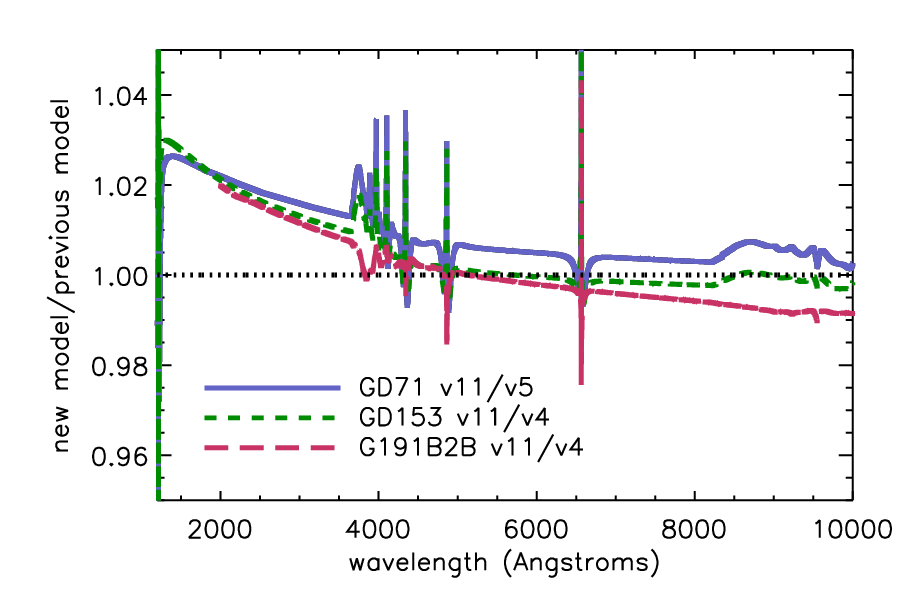}
    \caption{A comparison of the CALSPEC version 11 models of the three primary white dwarf standard stars compared to the older models that were used in the previous determination of the STIS L-mode sensitivities. The comparison is made in both the absolute flux (top) and the ratio (bottom).}
    \label{fig:WD_models}
\end{figure}

Figure \ref{fig:WD_models}  shows the previous and current CALSPEC models for the three standard stars over the wavelength range of STIS.  The first panel shows fluxes of the previous models in black and the current models in pink, green, and purple (for \GB, \GDone, and \GDseven, respectively). The most notable change is the inclusion of metal line absorption in \GB, which leads to strong absorption features in the UV below $\sim 2000$~\AA.  The second panel shows the ratios of the newer models over the previously used models, excluding the forest of absorption features in \GB\ for clarity.  The overall differences are small, ranging from $\sim3$\% increases at the shortest STIS wavelengths for all three stars to $\pm\sim$1\%  at the longest STIS wavelengths, depending on the standard star.  The average change over the three standard stars is near unity over much of the optical portion of the spectrum. 

\vspace{-0.3cm}
\ssection{STIS Observations}\label{sec:obs}

\begin{deluxetable}{lcccc}
    \tabcolsep 4pt
    \tablewidth{0pt}
    \tablecaption{Summary of observations used \label{tab:obs_sum}}
    \tabletypesize{\footnotesize}
    \tablehead{
      \colhead{Mode} & \colhead{Spectral range} & & \colhead{Number of observations} & \\
      \colhead{} & \colhead{\AA} & \colhead{GD71} & \colhead{GD153} &\colhead{G191B2B} }
    \startdata
    G140L & 1150--1730  & 28 & 23 & $\ldots$ \\
    G230L &  1570--3180 & 19 & 19 & $\ldots$ \\
    G230LB &  1680--3060 & 19 & 18 & 14 \\
    G430L &  2900--5700 & 26 & 24 & 19 \\
    G750L &   5240--10270 & 11 & 11 & 11\\
    \enddata
\end{deluxetable}

We used data from the full lifetime of STIS, utilizing programs that observed the primary WD standard stars to obtain high S/N for flux calibration purposes.  Some of these programs were designed to test sensitivities and calibration for the three STIS detectors, and observed a star at multiple positions across the detector.  A few other programs were designed to cross calibrate the Advanced Camera for Surveys (ACS) with STIS, and the more recent programs were part of a regular calibration monitoring campaign. The latter observed the three primary standard stars every two cycles. The white dwarfs \GDseven\ and \GDone\ were observed with all 5 L-modes, whereas \GB\ was observed only with the CCD L-modes (G230LB, G430L, and G750L).  \GB\ is too bright to safely be observed with the L-modes on the MAMA detectors. A summary of the available data for each mode is given in \autoref{tab:obs_sum}.
A more detailed list of all the observations from these programs is given in Appendix A Table \ref{tab:obs}, which provides the target name, observing mode, date and time of the observation, exposure time, observation ID, proposal ID, and any additional notes or flags.
All observations used the widest long slit (52x2$''$, the ``photometric'' slit). CCD observations were taken at both the nominal (middle of the detector) and E1 (near the top of the detector) positions, and the latter have ``E1'' marked in the notes column of  Table \ref{tab:obs}.  Most CCD data were taken with Gain=4 to access the full well depth of the CCD. Data taken at Gain=1 are noted with ``G1'' in the notes column. 

For G140L, the shadow of the repeller wire falls near the nominal position of spectra on the detector. Therefore, STIS's mode select mechanism by default offsets  the spectrum $\sim 3''$ from the detector center. Initially, this offset was towards the top of the detector. In March 1999, the offset was changed to move spectra to the lower part of the detector to avoid the FUV-MAMA glow region.  Typically, both G140L and G230L spectra are subject to additional monthly offsets to minimize wear on the detectors. However, special commanding was used to disable these offsets for calibration observations for better repeatability after March 1999. Any MAMA observations observed at non-standard positions on the detector are marked with ``P'' in the notes column of Table \ref{tab:obs}. 

\vspace{-0.3cm}
\ssection{Data Preparation and Sensitivity Derivation}\label{sec:sensitivity}
\ssubsection{Overview of Methodology}
The wavelength dependent sensitivity, $S_\lambda$, of a given STIS observing mode is the conversion factor between the incidence flux, $F_{\lambda}$, and the resultant count rate on the STIS detectors, $C_{\lambda}$. For a given observation, it is measured for a source with a known model flux, $F_{\lambda, model}$, as given in \autoref{eq_flux1}.
\begin{equation}
S_{\lambda,obs} = C_{\lambda,obs} / F_{\lambda,model}
\label{eq_flux1}
\end{equation} 
There are a number of factors that affect the sensitivity of a given observation with STIS that are corrected by the \calstis\ pipeline.  The flux calibration step in \calstis\  is performed in \textit{calstis6} (\href{https://www.stsci.edu/files/live/sites/www/files/home/hst/instrumentation/stis/documentation/instrument-science-reports/_documents/199903.pdf}{McGrath et~al.\ 1999}), following the equation in Section 3.4.13 in the STIS Data Handbook (DHB), and reproduced in Equation \ref{eq_flux2} below. 
\begin{equation}
F_{\lambda,calstis} = \left(\frac{1}{R_\lambda}\right)
\left(\frac{10^8\cdot h\cdot c\cdot H}{A_{HST} \cdot T_\lambda\cdot \lambda \cdot \Delta\lambda \cdot f_{\rm GAC}}\right)
\left(\frac{G}{f_{TDS}\cdot f_T}\right)
C_{\lambda, NET}
\label{eq_flux2}
\end{equation}
$F_{\lambda,calstis}$ and $C_{\lambda,NET}$ are the flux and count rate vectors in the one-dimensional spectra files produced by the \calstis\ pipeline (the ``FLUX'' and ``NET'' columns, respectively of the \texttt{\_x1d.fits} or \texttt{\_sx1.fits} files).
$R_\lambda$ is the wavelength dependent integrated throughput, which is stored in the \texttt{PHOTTAB} reference file for each optical element.  This is the quantity affected by the updates to the primary standard stars' model fluxes.
The terms in the second group of parentheses consists of physical constants, instrument constants, and correction factors contained in other reference files (see the STIS DHB for details). The terms in the third set of parentheses are the gain ($G$, CCD only) and correction factors for the time dependent sensitivity ($f_{\rm TDS}$) and the temperature dependent sensitivity ($f_{\rm T}$), which differ between observations. For CCD data, \textit{calstis6} also corrects the net count rates for charge transfer inefficiency (CTI)  prior to the flux calibration.

To determine $R_\lambda$, we must first correct the net count rate of each standard star observation for observation-specific sensitivity variations.  The average corrected net count rates for each primary star are then divided by the updated model fluxes to create an average sensitivity vector per standard star, $S_{\lambda,star}$.  Finally, $R_\lambda$ is calculated from averaging over the sensitivities and scaling by the telescope and instrument constants as in \autoref{eq_flux3} below.
\begin{equation}
R_\lambda = \overline{S_{\lambda,star}}
\left(\frac{A_{HST} \cdot T_\lambda\cdot \lambda \cdot \Delta\lambda \cdot f_{\rm GAC}}{10^8\cdot h\cdot c\cdot H}\right)
\label{eq_flux3}
\end{equation}

\begin{figure}[!p]
  \centering
  \includegraphics[scale=0.8,trim={1cm 0 0.8cm 0},clip]{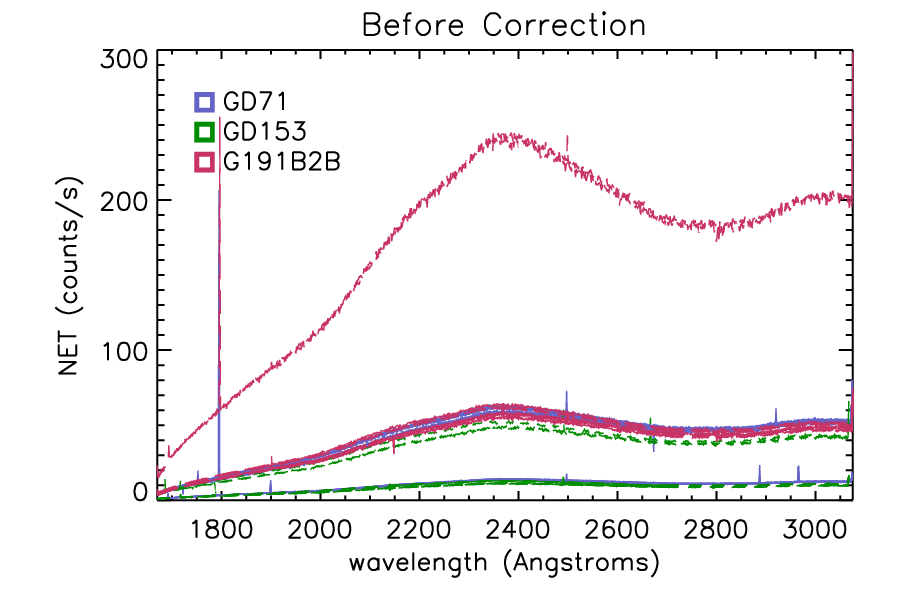}
    \includegraphics[scale=0.8,trim={1cm 0 0.8cm 0},clip]{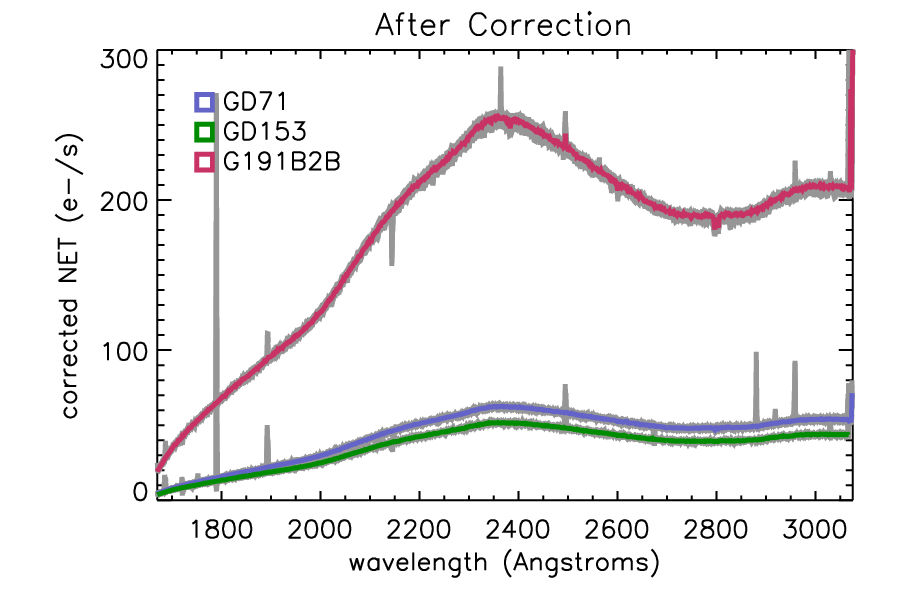}
    \caption{The top panel shows the raw NET count rates of individual G230LB observations of the three primary standards. In the second panel, the NET count rates have been multiplied by a correction factor to account for time and temperature sensitivity variations and converted to  electrons per second (taking into account the CCD gain).  The individual spectra are now shown in gray with the averages overlaid in color.  Most of the spread between observations of a given target is gone.}
    \label{fig:G230LB_counts}
\end{figure}

\ssubsection{Detailed Steps}
In this work, we used the net count rates in the extracted one-dimensional spectra files produced by the \calstis\ pipeline using the default extraction heights and background regions for each mode.  Using the default parameters ensures that the updated absolute flux calibration is optimized for the typical STIS data products in the MAST archive.  
The NET data have already been corrected for sky background and instrumental effects, such as flat field variation, bias and read noise, dark subtraction, and cosmic ray rejection. However, time ($f_{TDS}$) and temperature ($f_{T}$) dependent sensitivity changes have not yet been removed nor corrections for CTI for the CCD L-modes and the red halo correction for G750L (as in \autoref{eq_flux2} above).  To determine the cumulative effect of these corrections, we run \calstis\ twice -- the first with the usual defaults and corrections and the second with setting the primary header keywords \texttt{TDSTAB} to ``\texttt{N/A}'' and, for CCD modes, \texttt{CTECORR} to ``\texttt{OMIT}''.  The ratio of the FLUX columns of the resulting pairs of \texttt{\_x1d} files yields the correction factors for the observation-specific sensitivity variations. These correction factors are applied to the NET to correct them to a common STIS epoch and detector temperature.  In Figure \ref{fig:G230LB_counts}, we use G230LB observations as an example showing the NET count rates for each primary standard before and after applying these corrections and gain factor.

CCD observations are also impacted by cosmic rays.  By default, CCD exposures are  split into two identical sub-exposures (CRSPLIT=2) to allow for cosmic ray removal in \calstis.  Occasionally, telescope jitter during the observation can introduce small spatial offsets between the sub-exposures, leading the cosmic ray rejection algorithm to reject substantial valid data in the spectral extraction region as cosmic rays. When the two sub-exposures are combined, the resulting NET count rate will then be systematically low.  Following the recommendations in \href{https://ui.adsabs.harvard.edu/abs/2019stis.rept....2C/abstract}{Carlberg (2019)} and \href{https://www.stsci.edu/files/live/sites/www/files/home/hst/instrumentation/stis/documentation/instrument-science-reports/_documents/2021_01.pdf}{Hernandez (2021)}, we inspected the number of cosmic-ray-rejected pixels  in the spectral extraction region compared to the full CCD.  If the number rejected was higher than 20\%, we removed those data from the analysis. For GD71 with G750L, we lowered the threshold to 12\% to further remove obvious outliers and bad data.  Two G430L and several G750L observations were flagged, and they are marked in the Table \ref{tab:obs} as ``HCR'' (High Cosmic Rays).  We did attempt to recover  erroneously rejected counts by running \texttt{ocrreject} independently and increasing the threshold for cosmic ray rejection. However, this workflow alters the order of the cosmic ray rejection, dark subtraction, and flat fielding compared to running \calstis\ start to finish, and we found the results to be unreliable (see \href{https://hst-docs.stsci.edu/stisdhb/chapter-3-stis-calibration/3-5-recalibration-of-stis-data#id-3.5RecalibrationofSTISData-3.5.43.5.4ImprovingCosmicRayRejection}{STIS DHB section 3.5.4 ``Improving Cosmic Ray Rejection''} for more details).  Additionally there were a few observations from this calibration program that were not included because of poor data quality or non-standard observing modes (e.g., using sub-arrays).

G750L spectra are additionally affected by fringing at wavelengths $>7000$~\AA, due to interference of multiple reflections of long-wavelength light in the detector.   The  \textit{defringe} tool in the \texttt{stistools} package was used to correct fringes, using contemporaneously observed fringe flats.  We followed the same procedure described in \href{https://www.stsci.edu/files/live/sites/www/files/home/hst/instrumentation/stis/documentation/instrument-science-reports/_documents/2021_01.pdf}{Hernandez (2021)}, except when we ran \texttt{stistools.x1d.x1d()} after dividing by the fringe flat, we ran it twice, one with the usual defaults and one with the TDS and CTE corrections off, like for G230LB and G430L\footnote{Note, that at the time of the analysis, the standalone \texttt{stistools.x1d.x1d()} behavior was not altered by changing the primary header keyword calibration switches of the data files.  Instead one must set the corresponding arguments of the \texttt{stistools.x1d.x1d()} task itself, e.g.,  `CTECORR=OMIT' to alter the calibration.}.  We then found the corrected NET count rates from the ratio of the fluxes from the final defringed \texttt{\_x1d} products. 
\begin{figure}[!t]
  \centering
  \includegraphics[scale=0.7]{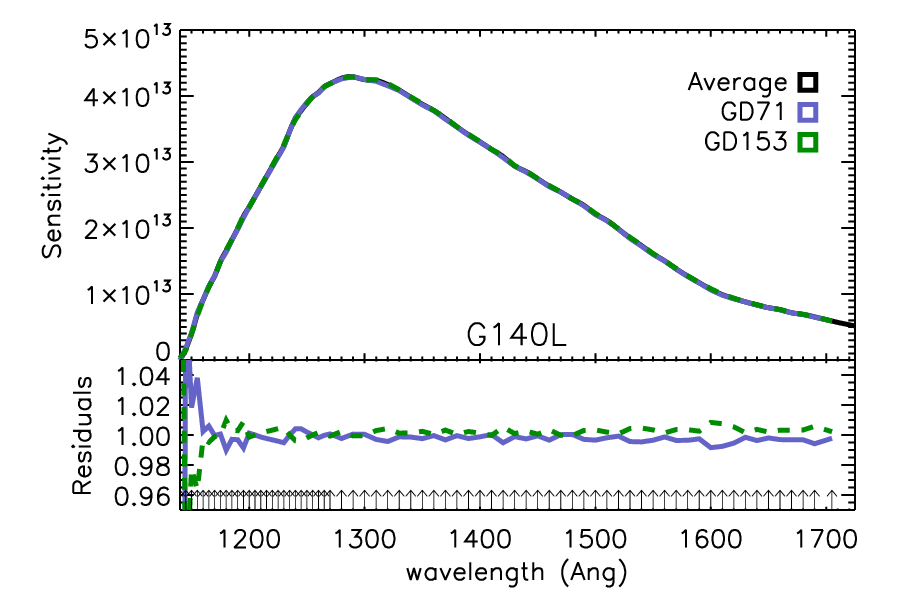}
  \includegraphics[scale=0.7]{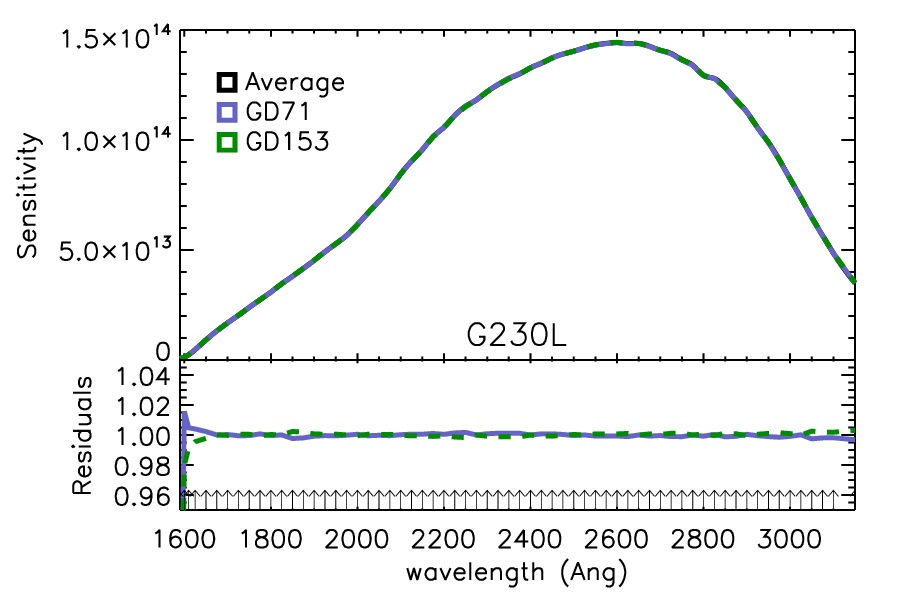}
    \caption{The sensitivities and residuals for each standard star and the average sensitivity for G140L (top) and G230L (bottom).  The spline nodes are overlaid at the bottom of each as arrows.}
    \label{fig:G140L_sens}
\end{figure}

Once the corrected NET count rates are in hand, the data are placed on a common wavelength scale using a flux conserving interpolation. A heliocentric correction was not applied to each spectrum since this correction is about 100 times smaller than the velocity resolution of the L-modes.  All data for a given standard star are then combined with a sigma-clipped mean at each wavelength pixel. This step has the added benefit of removing uncorrected hot pixels and/or cosmic rays missed by the \texttt{DARKCORR} and \texttt{CRCORR} steps of \calstis\ (e.g., the gray spikes in the bottom panel of Figure \ref{fig:G230LB_counts}).

\begin{figure}[!ht]
  \centering
  \includegraphics[scale=0.6]{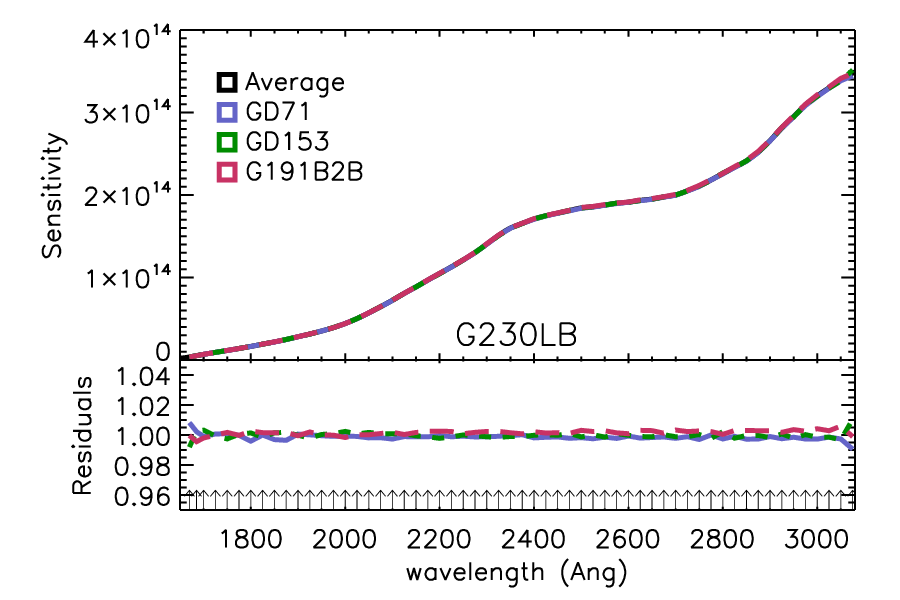}
  \includegraphics[scale=0.6]{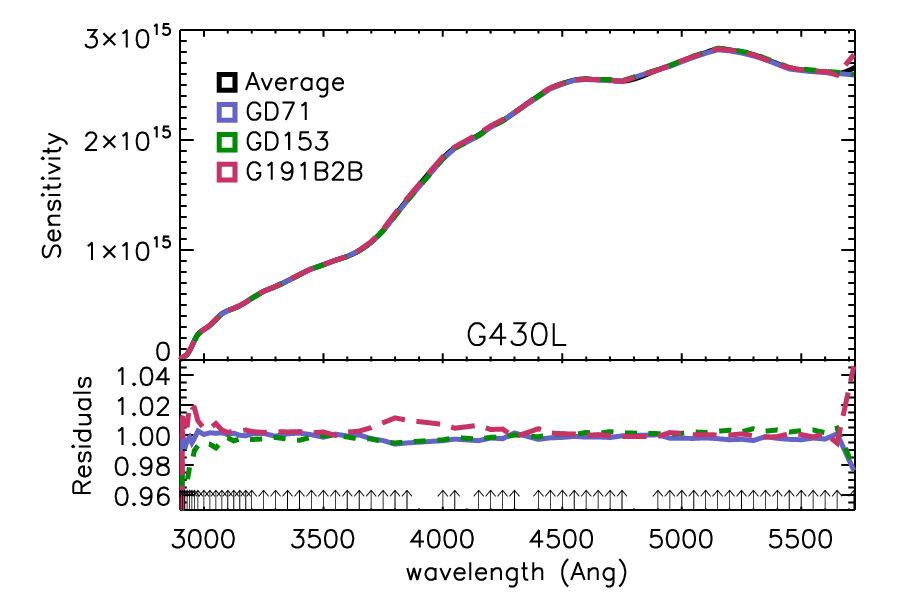}
  \includegraphics[scale=0.6]{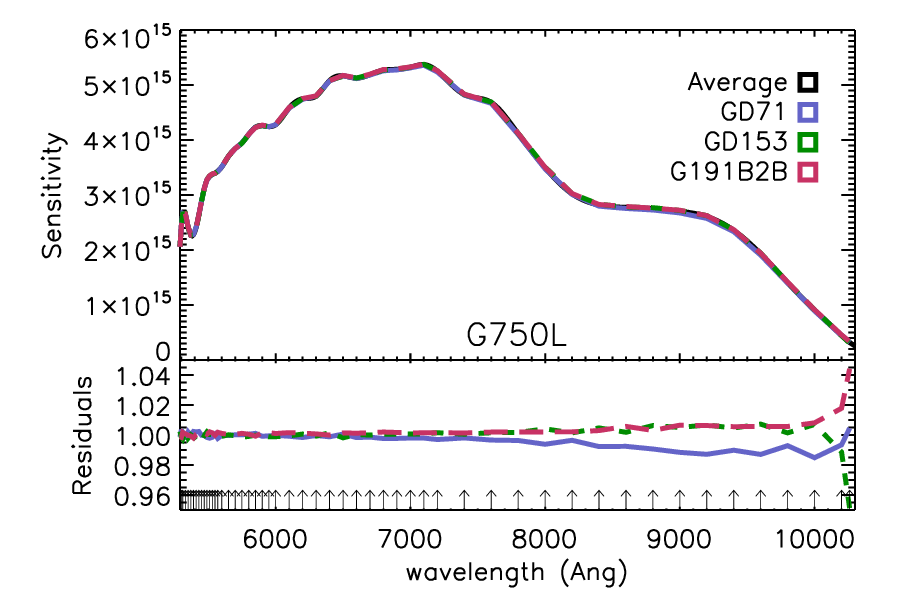}
    \caption{Same as \autoref{fig:G140L_sens} for the 3 CCD gratings.}
    \label{fig:G140L_sens2}
\end{figure}
\clearpage

The sensitivity of each mode ($S_{\lambda, star}$) is first measured for each star by dividing the average corrected NET count rates by the corresponding CALSPEC flux model, which has been similarly interpolated onto the same wavelength grid as the observations. These initial sensitivities are fit with a customized spline for each mode to ensure the curves are finely sampled enough to capture any changes in the sensitivities or model fluxes.  The spline nodes are shown as arrows at the bottom of each panel in \autoref{fig:G140L_sens} and \autoref{fig:G140L_sens2}.  Emission or absorption features are masked from the fit, including both the cores and wings of the Hydrogen Lyman and Balmer series lines. This is because the models still do not reproduce these features well. The spline fits for each available primary standard are then averaged together with equal weights to obtain the final sensitivity per observing mode ($\overline{S_{\lambda, star}}$).  The sensitivities for each star and the average per mode are shown in \autoref{fig:G140L_sens} and \autoref{fig:G140L_sens2}.  The difference between the star's sensitivity and the average is also shown in the bottom panel for each mode, labeled as the `Residuals'.  The residuals are 
typically better than 1\% over most of the spectral range of each grating, though it rises to $\sim$2\% at the edges. The exception is very red end of G750L where residuals increase to about 4\%.  See \autoref{sec:G750L} for more details about G750L.

These average sensitivities per mode, in units of (counts/s)/(ergs/cm$^2$/s/\AA) are then used to compute integrated throughputs ($R_\lambda$) for the \texttt{PHOTTAB} reference files used by \calstis. The sensitivities describe the telescope throughput as measured through the 52x2$''$ aperture, which is the mode used in the observations, and extracted using the nominal 7 or 11 pixel extraction box height, depending on the mode. They are then converted to throughput curves, defined as the efficiencies for an infinite aperture and infinite extraction box following the equation in \href{https://ui.adsabs.harvard.edu/abs/2012stis.rept....1B/abstract}{Bostroem et al.\ (2012)}.  The conversion also takes into account the collecting area of HST.  The \texttt{PHOTTAB} is then updated with the new throughputs for these observing modes (see \autoref{eq_flux3}).

\vspace{-0.3cm}
\ssection{Special Considerations for G750L}\label{sec:G750L}
Updated throughputs for G750L were delivered a year after those for the other four L-mode gratings.  There are several factors that complicate the G750L wavelength range, causing a compounding degradation of data quality at longer wavelengths. Already mentioned above are the fringes 
at wavelengths longer than 7000 \AA.  Additionally, white dwarfs are fainter at longer wavelengths leading to both lower signal-to-noise (S/N) per observation and longer exposure times, which increase the impact of cosmic rays.  In \autoref{tab:obs}, observations flagged with `HCR' occur disproportionally more often for the G750L observations.  Only 11 observations of each star were used in our analysis, the least of all the L-modes \autoref{tab:obs_sum}.

The point spread function (PSF) of STIS also has a strong wavelength dependence for G750L, where the enclosed energy for the default 7-pixel extraction height drops below $\sim80$\% redward of $\sim 8000$~\AA\ (\href{https://www.stsci.edu/files/live/sites/www/files/home/hst/instrumentation/stis/documentation/instrument-science-reports/_documents/199713.pdf}{Leitherer \& Bohlin 1997}). The excess flux in the wings of the PSF at long wavelengths is referred to as the ``red halo''. With fractionally more light in the PSF wings, the 
photometric precision at the longest wavelengths is much more sensitive to focus variations (see, e.g., \href{https://ui.adsabs.harvard.edu/abs/2017stis.rept....1P/abstract}{Proffitt et~al., 2017}). The CTE correction for G750L also accounts for the red halo (see the \href{https://hst-docs.stsci.edu/stisdhb/chapter-3-stis-calibration/3-4-descriptions-of-calibration-steps#id-3.4DescriptionsofCalibrationSteps-3.4.63.4.6CTECORR:CorrectionforChargeTransferInefficiencyLosses}{STIS DHB Chapter 3.4.6 `CTECORR'} and \href{https://www.stsci.edu/files/live/sites/www/files/home/hst/instrumentation/stis/documentation/instrument-science-reports/_documents/200603.pdf}{Goudfrooij 2006}) to avoid underestimating the number of charge traps filled by electrons outside the extraction box. 
To minimize these complications, the custom pipeline used by CALSPEC adopts an 11-pixel extraction height for G750L.  
The \calstis\ pipeline which is the default pipeline, in contrast, adopts a 7 pixel extraction height for all CCD data as an optimal trade-off between achievable S/N and photometric precision for typical moderate S/N STIS spectra (\href{https://www.stsci.edu/files/live/sites/www/files/home/hst/instrumentation/stis/documentation/instrument-science-reports/_documents/199713.pdf}{Leitherer \& Bohlin 1997}).  Since our sensitivities will be used by \calstis\, we also maintain a 7 pixel extraction height for consistency.  
However, we tested a number of variations in how the G750L spectra were extracted prior to the sensitivity derivation, including increasing the extraction height to 11 pixels and re-factoring the CTE correction  so that the CTE reference time (2002) would match that used by TDS (1997).  None of these variations resulted in significantly different sensitivity shapes. Therefore, we conclude that our method of averaging the sensitivity derivations per standard star and averaging again over all three standard stars is robust to the interplay of fringing, lower S/N, and increased PSF width affecting these wavelengths that lead to the larger residual scatter in  \autoref{fig:G140L_sens2}.

\vspace{-0.3cm}
\ssection{Validation of Updated Throughputs}\label{sec:testing}

\begin{figure}[!ht]
  \centering
  \includegraphics[scale=0.7]{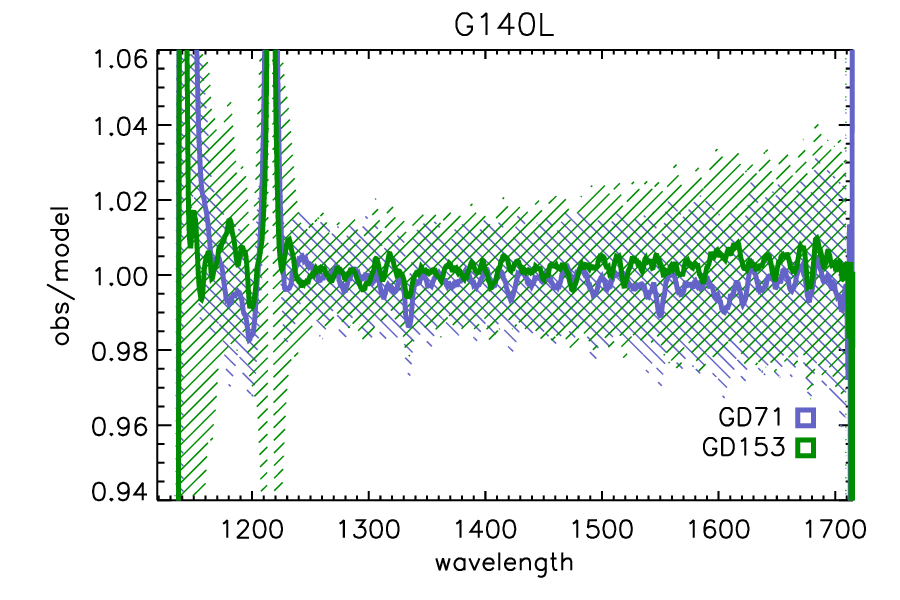}
  \includegraphics[scale=0.7]{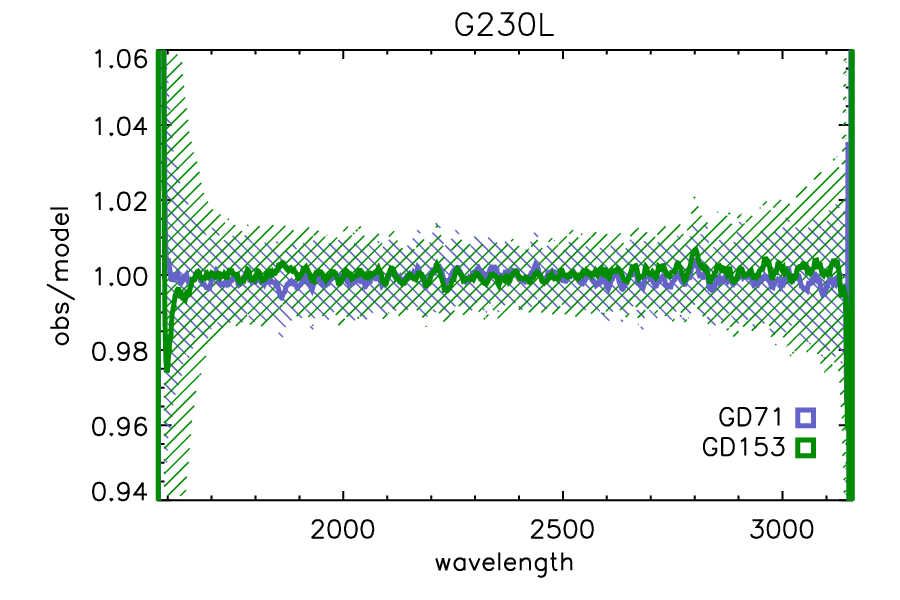}
    \caption{Plots showing the ratio of the smoothed average flux of the data set after being re-run with the updated \texttt{PHOTTAB} over the CALSPEC model for each standard star for the MAMA detectors.  The standard deviations are shown as the banded regions. Typically, the re-reduced data agree with the model to better than 1\% on average, and within 2\% for individual datasets. Exceptions include the grating edges, where S/N is lowest, and in the cores of narrow absorption or emission features (e.g., the Lyman $\alpha$ line in G140L).}
    \label{fig:verify}
\end{figure}

\begin{figure}[!ht]
  \centering

  \includegraphics[scale=0.6]{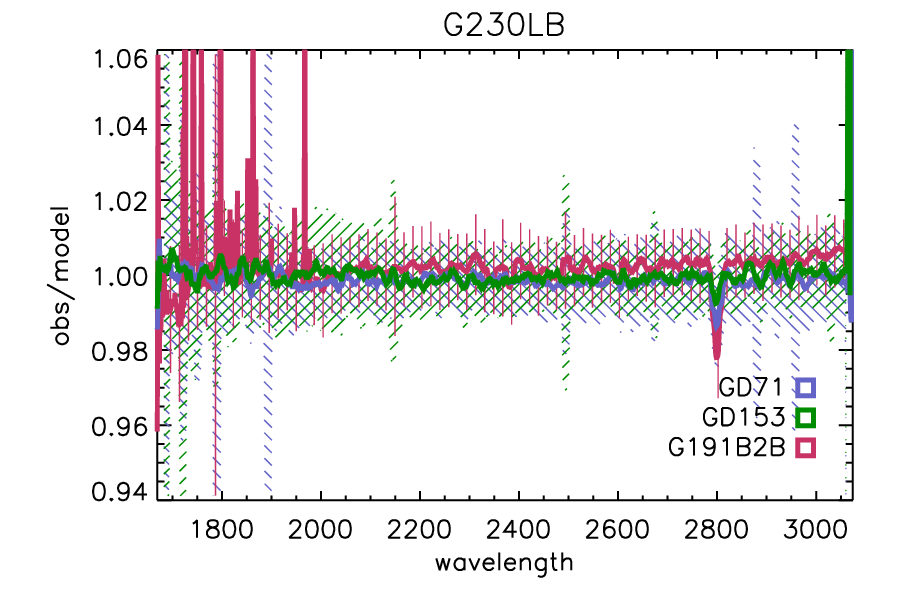}
  \includegraphics[scale=0.6]{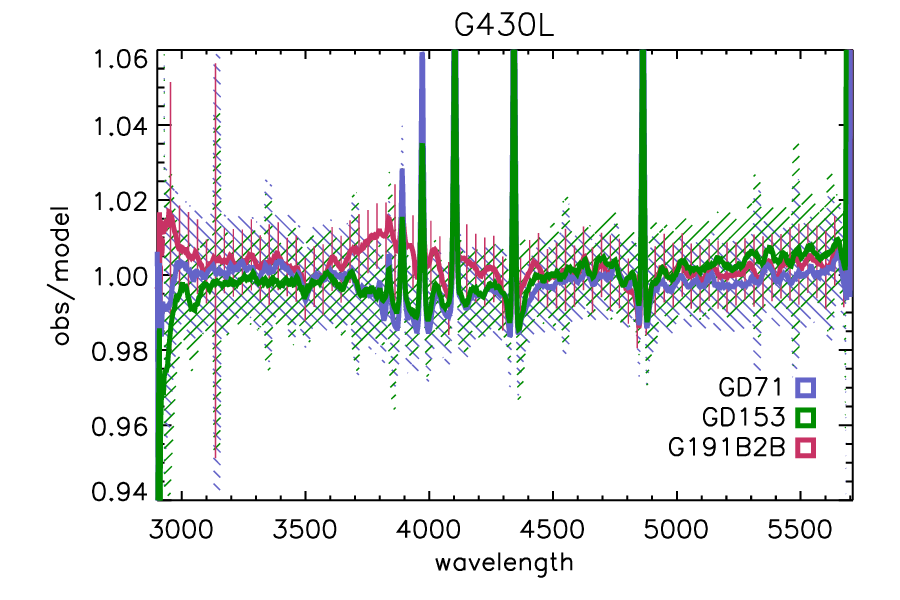}
  \includegraphics[scale=0.6]{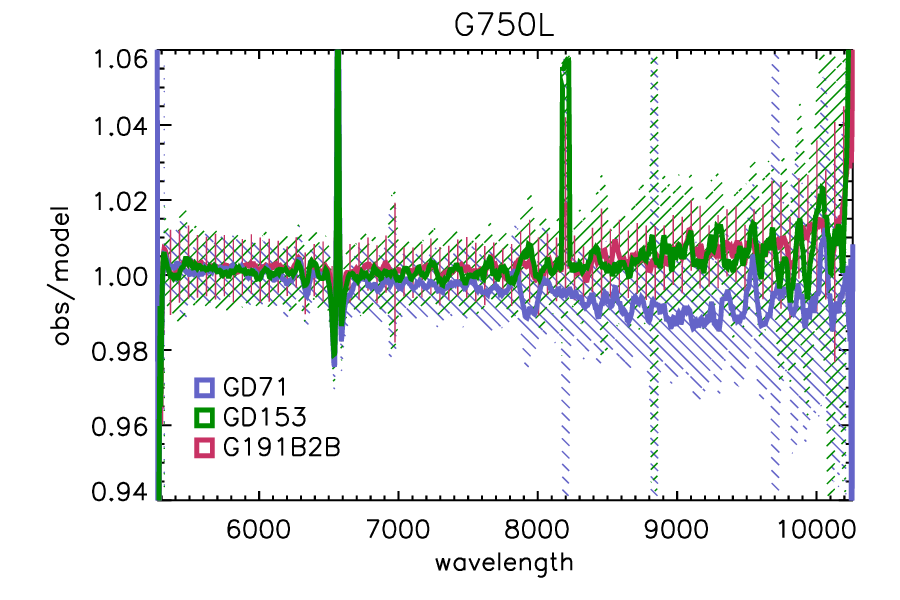}
    \caption{Same as Figure \ref{fig:verify} for the 3 CCD gratings. In the optical and near infrared, the broadband discrepancies at the $\sim$1.5\% level between the models are likely due to limitations in the modeling. In G430L data, \GB\ deviates from other two standards between the Balmer jump ($\sim3647$~\AA) and Balmer $\delta$ ($\sim4341$~\AA). Similar discrepancies arise for G750L data of \GDseven, from the Paschen jump ($\sim8206$~\AA) to beyond the Paschen $\gamma$ line ($\sim10052$~\AA) }
    \label{fig:verify2}
\end{figure}

We verified the new sensitivities in a few ways.  After creating a new \texttt{PHOTTAB} based on the updated sensitivities for each mode, we re-reduced all the data for the primary standard stars through \calstis\  with the new \texttt{PHOTTAB} (including defringing G750L spectra) to compare with the CALSPEC models. The results are shown in Figures \ref{fig:verify} and \ref{fig:verify2} for each L-mode as the re-processed data divided by the models.  The fluxes were median smoothed with a 25 pixel kernel. The average spectrum per star per mode is less than 1\% different from the model at all wavelengths, except at edges where there is less throughput and signal, at wavelengths $>$8000\AA\ (see \autoref{sec:G750L}), and at emission/absorption features where we expect differences between the data and models (see \href{https://ui.adsabs.harvard.edu/abs/2020AJ....160...21B/abstract}{Bohlin et al 2020}).  The standard deviations are shown as the banded regions and are typically less than 2\%.  Sharp spikes seen only in the banded region (i.e., very local high standard deviation) typically show up only for the CCD gratings (G230LB, G430L, and G750L) and are likely due to unstable hot pixels that are not optimally removed by the monthly dark file. 

In Figures \ref{fig:verify} and \ref{fig:verify2}, the sharp spikes in the averages are typically discrepancies in either the narrow metal absorption features of \GB\ (below $\sim$2000~\AA\ in G230LB, which show up because the models have not been convolved to the STIS spectral resolution) or in the cores of the strong hydrogen absorption features. These include Lyman $\alpha$ in the G140L passband, the Balmer series (with the exception of Balmer $\alpha$) in the G430L passband, and Balmer $\alpha$ and Paschen $\delta$ and higher energy lines in G750L.  Broadband discrepancies between the different standard stars arise at the Balmer jump ($\sim3647$~\AA) and the Paschen bump ($\sim8206$~\AA), highlighting the limitations of the atmospheric modeling. \GB\ is the outlier standard star at the Balmer jump, while \GDseven\ is the outlier at the Paschen jump. 

\clearpage
\clearpage

\begin{figure}[!ht]
  \centering
  \includegraphics[scale=0.7]{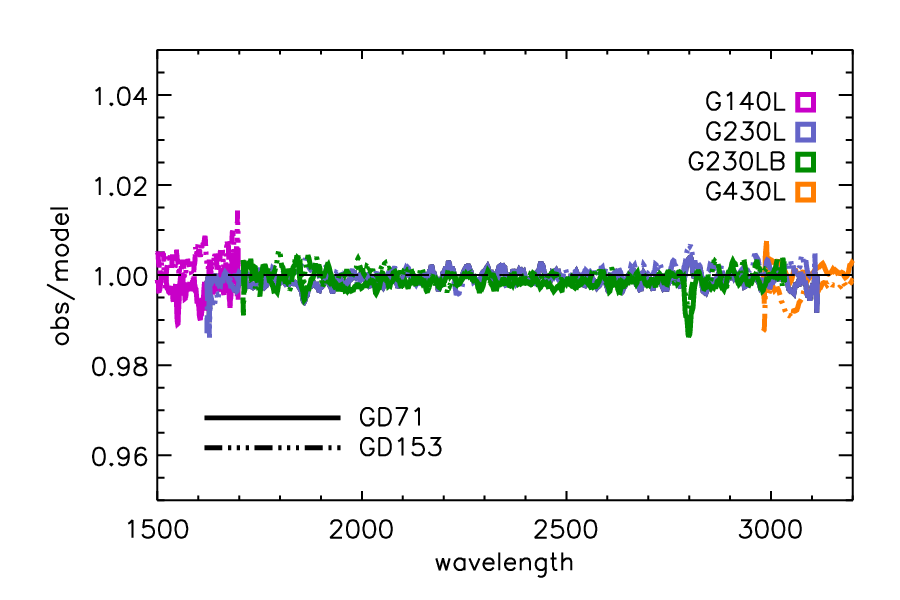}
  \includegraphics[scale=0.7]{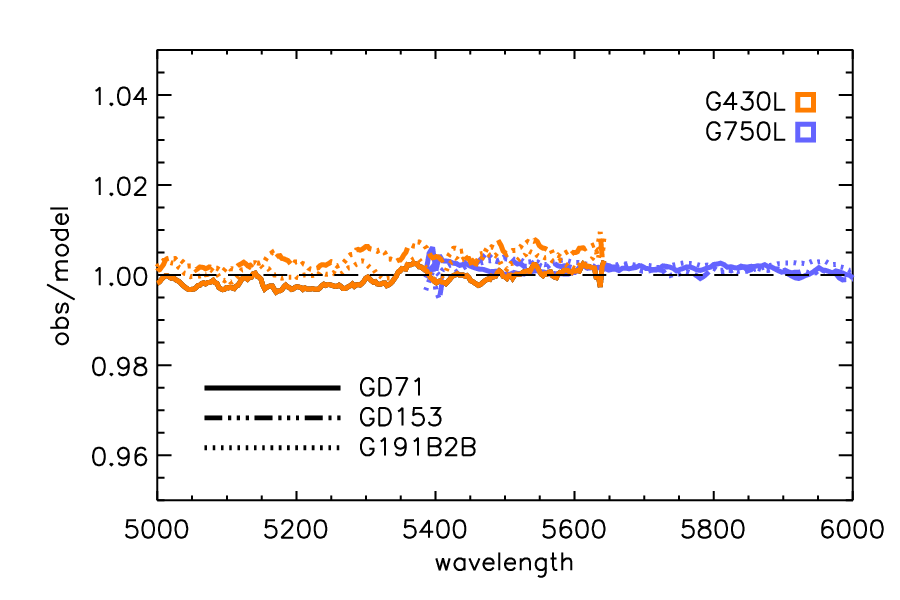}
    \caption{The ratio of the smoothed average flux of the data after re-reducing it with the updated \texttt{PHOTTAB} to the CALSPEC model in wavelength regions where some of the L-mode gratings overlap.  The different gratings are shown as the different colors with the standard stars as different line styles.  For clarity, the pixels at the edges were removed.  There are no systematic differences seen between the different gratings.}
    \label{fig:overlap}
\end{figure}

We also looked at the regions where the wavelength ranges of some of the L-modes overlap.  G230L and G230LB cover a similar wavelength range in the NUV and overlap on the ends with G140L and G430L.  G430L and G750L also overlap.  These overlapping regions are shown in \autoref{fig:overlap}.  For clarity, the pixels at the edges were removed since those tend to fluctuate more due to the lower S/N.  The average ratio of the fluxes from the re-processed data and the model are shown in different colors for the different gratings, and the stars are plotted with different line styles.  There is good agreement between the L-modes, with differences usually less than 1\%.

\begin{figure}[!ht]
  \centering
  \includegraphics[scale=0.47]{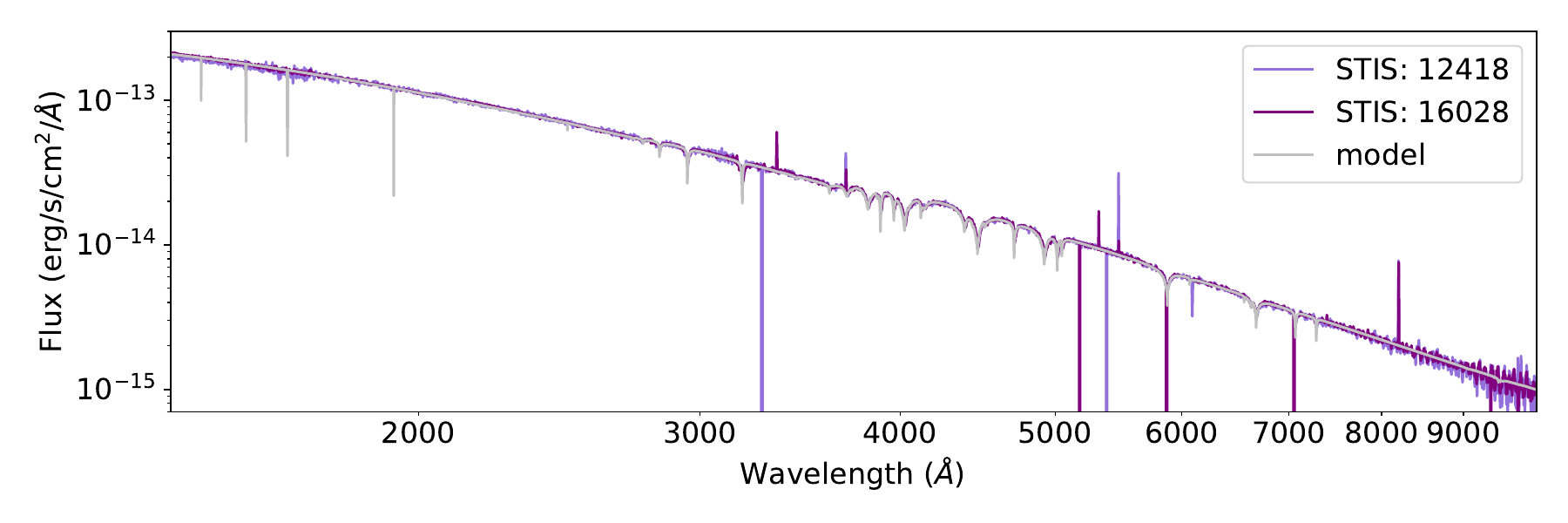}
  \includegraphics[scale=0.47]{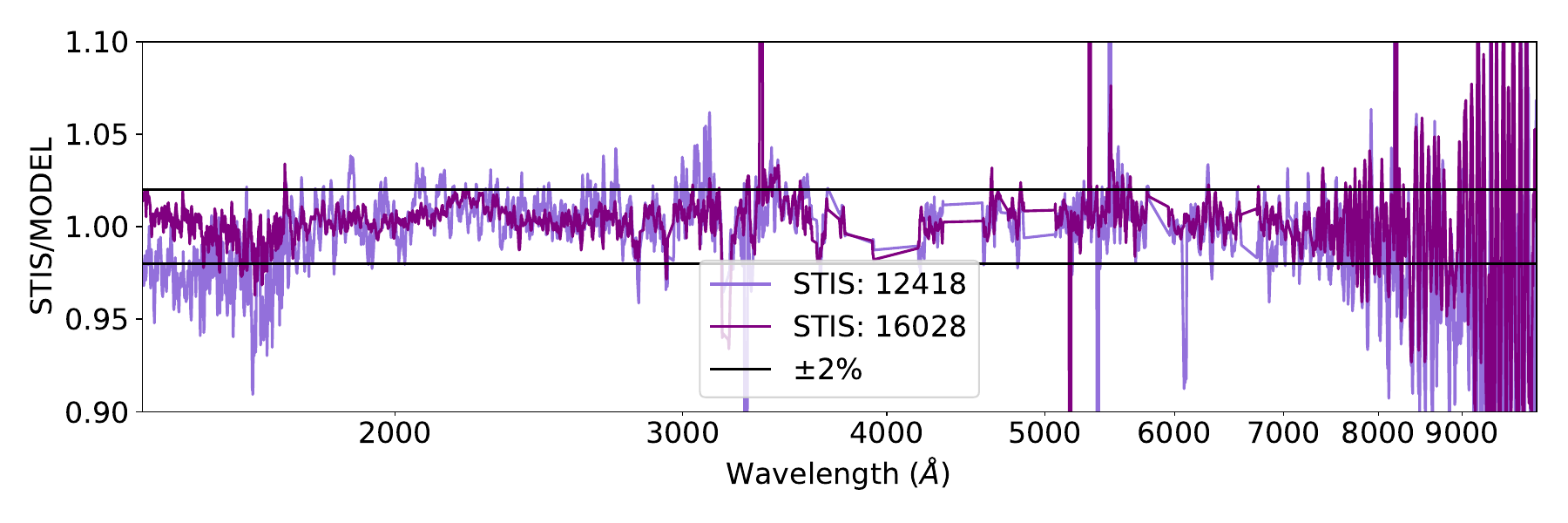}
    \caption{(Top): STIS observations (purple lines) of COS standard star WD~0308-565 from the FUV through near IR compared to the CALSPEC model atmosphere (gray). (Bottom): The same data plotted as a ratio relative to the model, with residuals in deep absorption features removed and the ratio spectrum smoothed with a boxcar filter. The data are HASP co-added spectra from two separate HST observing programs. The long wavelength data have not been defringed. The median S/N for data from program 16028 (dark purple) is 96, while for program 12418 (light purple), the median S/N is only 50.
   }
    \label{fig:overlap2}
\end{figure}

Lastly, for a check independent of the data we used to re-derive the sensitivity, we compare \calstis-reduced STIS data of the helium white dwarf WD~0308-565, which is used as a flux standard for the HST/COS instrument (e.g., \href{https://www.stsci.edu/files/live/sites/www/files/home/hst/instrumentation/cos/documentation/instrument-science-reports-isrs/_documents/COS_ISR_2024_16.pdf}{Miller et~al, 2024}), to its CALSPEC model. For ease of comparison across the full spectral range of STIS, we retrieved the Hubble Advanced Spectral Product (HASP, \href{https://ui.adsabs.harvard.edu/abs/2024cos..rept....1D/abstract}{Debes et~al., 2024}) program-level co-added spectra from program IDs 12418 and 16028. Figure \ref{fig:overlap2} shows both the flux comparison of the data compared to the CALSPEC model and the ratio of the data to the model as a function of wavelength for the full wavelength range covered by the five STIS L-mode gratings. The HASP spectral products do not correct for fringing in G750L, which leads to much larger residuals at the longest wavelengths. In the top panel of Figure \ref{fig:overlap2}, the model spectra have not been convolved to the (lower) resolution of the STIS spectra, so the cores of deep spectral lines or lines with intrinsically narrow widths show apparent discrepancies with respect to the observations. However, the shapes of the lines with large intrinsic broadening (e.g., between $\sim$3800--5000~\AA) are well matched between the models and observations.

In the ratio panel of Figure \ref{fig:overlap2}, strong spectral features are masked, and the ratio  has  been smoothed by a 5 point boxcar filter for visual clarity. Horizontal lines at $\pm2$\% have been plotted to guide the eye. The individual spectra show variations consistent with what was seen in Figures \ref{fig:verify} and \ref{fig:verify2}. It is noteworthy that the regions of largest discrepancy coincide with the lowest S/N, particularly at the longest and shortest wavelengths where there is no overlapping wavelength coverage to compensate for the low sensitivity at the grating edges (Figure \ref{fig:G140L_sens}).


\vspace{-0.3cm}
\ssectionstar{Acknowledgments}
\vspace{-0.3cm}
The authors are incredibly grateful to Ralph Bohlin for his support, expertise and tireless efforts to maximize STIS's calibration accuracy. We also thank Doug Long for supporting the early stages of this initiative. 

\vspace{-0.3cm}
\ssectionstar{Change History for STIS ISR 2025-02}\label{sec:History}
\vspace{-0.3cm}
Version 1: \ddmonthyyyy{30 July 2025} - Original Document 

\vspace{-0.3cm}
\ssectionstar{References}\label{sec:References}
\vspace{-0.3cm}

\noindent
\href{https://www.stsci.edu/files/live/sites/www/files/home/events/event-assets/2002/_documents/2002-hst-calibration-workshop-presentation-bohlin.pdf}{Bohlin, R.,  2002, \textit{HST Calibration Workshop}, ed. S. Arribas, A. Koekemoer, \& B. Whitmore (Baltimore: STScI), 115} \\
\href{https://ui.adsabs.harvard.edu/abs/1995AJ....110.1316B/abstract}{Bohlin, R.C., Colina, L., \& Finley, D.S. 1995, AJ, 110, 1316} \\
\href{https://ui.adsabs.harvard.edu/abs/2019AJ....158..211B/abstract}{Bohlin, R.~C., Deustua, S.~E., \& de Rosa, G.\ 2019, AJ, 158, 211} \\
\href{https://ui.adsabs.harvard.edu/abs/2004AJ....128.3053B/abstract}{Bohlin, R.~C., Gilliland, R.~L., 2004, AJ, 128, 6} \\
\href{https://ui.adsabs.harvard.edu/abs/2014PASP..126..711B/abstract}{Bohlin, R.~C., Gordon, K.~D., \& Tremblay, P.-E.\ 2014, PASP, 126, 711} \\
\href{https://ui.adsabs.harvard.edu/abs/2020AJ....160...21B/abstract}{Bohlin, R., Hubeny, I., \& Rauch, T. 2020, AJ, 160, 21}\\
\href{https://ui.adsabs.harvard.edu/abs/2012stis.rept....1B/abstract}{Bostroem, K. A., Aloisi, A., Bohlin, R., Hodge, P., \& Proffitt, C. 2012, STIS Instrument Science Report 2012-01}\\
\href{https://ui.adsabs.harvard.edu/abs/2019stis.rept....2C/abstract}{Carlberg, J., 2019, STIS Instrument Science Report 2019-02}\\
\href{https://ui.adsabs.harvard.edu/abs/2022stis.rept....4C/abstract}{Carlberg, J., et~al., 2022, STIS Instrument Science Report 2022-4} \\
\href{https://ui.adsabs.harvard.edu/abs/2024cos..rept....1D/abstract}{Debes, J., et~al., 2024, COS Instrument Science Report 2024-01} \\
\href{https://www.stsci.edu/files/live/sites/www/files/home/hst/instrumentation/stis/documentation/instrument-science-reports/_documents/200603.pdf}{Goudfrooij, P., 2006, STIS Instrument Science Report 2006-03}\\
\href{https://www.stsci.edu/files/live/sites/www/files/home/hst/instrumentation/stis/documentation/instrument-science-reports/_documents/2021_01.pdf}{Hernandez, S., 2021, STIS Instrument Science Report 2021-01}\\
\href{https://ui.adsabs.harvard.edu/abs/2024stis.rept....4H/abstract}{Hernandez, S., et~al., 2024, STIS Instrument Science Report 2024-04} \\
\href{https://www.stsci.edu/files/live/sites/www/files/home/hst/instrumentation/stis/documentation/instrument-science-reports/_documents/199713.pdf}{Leitherer, C.\ \&  Bohlin, R., 1997, STIS Instrument Science Report 1997-13} \\
\href{https://www.stsci.edu/files/live/sites/www/files/home/hst/instrumentation/stis/documentation/instrument-science-reports/_documents/199903.pdf}{McGrath, M.~A., Busko, I, \& Hodge, P. 1999, STIS Instrument Science Report 1999-03} \\
\href{https://www.stsci.edu/files/live/sites/www/files/home/hst/instrumentation/cos/documentation/instrument-science-reports-isrs/_documents/COS_ISR_2024_16.pdf}{Miller, L., et~al., 2024, COS Instrument Science Report 2024-16} \\
\href{https://ui.adsabs.harvard.edu/abs/2017stis.rept....1P/abstract}{Proffitt, C.~R., Monroe, T., \& Dressel, L., STIS Instrument Science Report 2017-01} \\
\href{https://ui.adsabs.harvard.edu/abs/2013A%26A...560A.106R/abstract}{Rauch, T., Werner, K., Bohlin, R., \& Kruk, J. W. 2013, A\&A, 560, A106}\\
\href{https://hst-docs.stsci.edu/stisdhb}{Rickman, E., Brown J., et al., 2024, ``STIS Data Handbook'', Version 8.0, (Baltimore: STScI)} \\
\href{https://ui.adsabs.harvard.edu/abs/2024stis.rept....2S/abstract}{Siebert, M., et~al., STIS Instrument Science Report 2024-02}
\vspace{-0.3cm}

\ssectionstar{Appendix A: Observations}\label{sec:Appendix}
\vspace{-0.3cm}
\begin{deluxetable}{llccrcrl}
    \tabcolsep 4pt
    \tablewidth{0pt}
    \tablecaption{STIS Observational Data\label{tab:obs}}
    \tabletypesize{\footnotesize}
    \tablehead{
      \colhead{Target} &  \colhead{Mode} & \colhead{Date} & \colhead{Time} & \colhead{Exptime} & \colhead{Obs ID} &\colhead{Prop ID} & \colhead{Flags:} \\ 
      \colhead{} &  \colhead{} & \colhead{(YY-MM-DD)} & \colhead{(HH:MM:SS UT)} & \colhead{(s)} & \colhead{} &\colhead{} & \colhead{ \textbf{G1} Gain=1, \textbf{P} postarg used} \\ 
      \colhead{} &  \colhead{} & \colhead{} & \colhead{} & \colhead{} & \colhead{} &\colhead{} & \colhead{ \textbf{E1} 52x2E1 Aperture,} \\ 
      \colhead{} &  \colhead{} & \colhead{} & \colhead{} & \colhead{} & \colhead{} &\colhead{} & \colhead{\textbf{HCR} high CR rejection}
      }
    \startdata
    
G191B2B & G750L & 98-02-11 & 06:04:24 & 1980.0 & o49x07010 & 7674 &  HCR \\
G191B2B & G750L & 98-02-26 & 02:21:52 & 1980.0 & o49x08010 & 7674 &   HCR \\
G191B2B & G230LB & 97-10-18 & 18:27:14 & 150.0 & o4d101010 & 7805 &  \\
G191B2B & G430L & 97-10-18 & 18:35:59 & 150.0 & o4d101020 & 7805 &  \\
G191B2B & G750L & 97-10-18 & 18:44:44 & 1020.0 & o4d101030 & 7805 &  \\
G191B2B & G230LB & 97-11-22 & 19:10:38 & 150.0 & o4d102010 & 7805 &  \\
G191B2B & G430L & 97-11-22 & 19:19:23 & 150.0 & o4d102020 & 7805 &  \\
G191B2B & G750L & 97-11-22 & 19:28:08 & 1020.0 & o4d102030 & 7805 &  \\
G191B2B & G230LB & 01-01-20 & 06:07:25 & 240.0 & o69u05010 & 8849 &  \\
G191B2B & G430L & 01-01-20 & 06:17:46 & 240.0 & o69u05020 & 8849 &  \\
G191B2B & G750L & 01-01-20 & 06:28:07 & 1200.0 & o69u05030 & 8849 &  \\
G191B2B & G230LB & 01-02-22 & 11:18:11 & 240.0 & o69u06010 & 8849 &  \\
G191B2B & G430L & 01-02-22 & 11:28:32 & 240.0 & o69u06020 & 8849 &  \\
G191B2B & G750L & 01-02-22 & 11:38:53 & 1200.0 & o69u06030 & 8849 &  \\
G191B2B & G430L & 03-10-29 & 02:59:11 & 70.0 & o8v203020 & 10039 &  G1 \\
G191B2B & G430L & 03-10-29 & 03:49:11 & 70.0 & o8v203030 & 10039 & E1, G1 \\
G191B2B & G230LB & 03-10-29 & 03:56:42 & 150.0 & o8v203040 & 10039 & E1, G1 \\
G191B2B & G230LB & 03-10-29 & 04:01:22 & 150.0 & o8v203050 & 10039 &  G1 \\
G191B2B & G750L & 03-10-29 & 04:10:13 & 290.0 & o8v203060 & 10039 &  G1 \\
G191B2B & G430L & 10-03-15 & 13:03:29 & 200.0 & obbc07010 & 11889 & E1 \\
G191B2B & G430L & 10-03-15 & 13:08:59 & 200.0 & obbc07020 & 11889 &  \\
G191B2B & G230LB & 10-03-15 & 13:15:55 & 200.0 & obbc07030 & 11889 &  \\
G191B2B & G750L & 10-03-15 & 13:25:35 & 1310.0 & obbc07040 & 11889 &  HCR \\
G191B2B & G430L & 11-01-31 & 04:33:50 & 200.0 & obnf05010 & 12392 & E1 \\
G191B2B & G430L & 11-01-31 & 04:39:20 & 200.0 & obnf05020 & 12392 &  \\
G191B2B & G230LB & 11-01-31 & 04:46:15 & 200.0 & obnf05030 & 12392 &  \\
G191B2B & G750L & 11-01-31 & 04:55:56 & 1308.0 & obnf05040 & 12392 &  \\
G191B2B & G430L & 11-11-06 & 12:47:55 & 200.0 & obvp07010 & 12737 &  \\
G191B2B & G430L & 11-11-06 & 12:53:25 & 200.0 & obvp07020 & 12737 & E1 \\
G191B2B & G230LB & 11-11-06 & 13:00:20 & 216.0 & obvp07030 & 12737 & E1 \\
G191B2B & G750L & 11-11-06 & 13:10:17 & 1308.0 & obvp07040 & 12737 &  \\
G191B2B & G230LB & 12-10-14 & 18:00:31 & 468.0 & oc3i14010 & 12813 & E1 \\
G191B2B & G430L & 12-10-14 & 18:11:54 & 220.0 & oc3i14020 & 12813 & E1 \\
G191B2B & G430L & 12-10-14 & 19:01:02 & 220.0 & oc3i14030 & 12813 &  \\
G191B2B & G750L & 12-10-14 & 19:08:17 & 1100.0 & oc3i14040 & 12813 &  HCR \\
G191B2B & G230LB & 14-02-02 & 06:13:26 & 468.0 & ocga06010 & 13599 &  \\
G191B2B & G430L & 14-02-02 & 06:24:49 & 220.0 & ocga06020 & 13599 &  \\
G191B2B & G430L & 14-02-02 & 06:30:39 & 220.0 & ocga06030 & 13599 & E1 \\
G191B2B & G750L & 14-02-02 & 06:38:37 & 1058.0 & ocga06040 & 13599 &  \\
G191B2B & G230LB & 16-12-07 & 06:02:40 & 466.0 & odck03010 & 14861 &  \\
G191B2B & G430L & 16-12-07 & 06:14:01 & 216.0 & odck03020 & 14861 &  \\
G191B2B & G430L & 16-12-07 & 06:19:47 & 216.0 & odck03030 & 14861 & E1 \\
G191B2B & G750L & 16-12-07 & 07:32:53 & 1042.0 & odck03040 & 14861 &  \\
G191B2B & G230LB & 18-11-27 & 18:51:16 & 462.0 & odud03010 & 15602 &  \\
G191B2B & G430L & 18-11-27 & 19:02:33 & 214.0 & odud03020 & 15602 &  \\
G191B2B & G430L & 18-11-27 & 19:08:17 & 214.0 & odud03030 & 15602 & E1, HCR \\
G191B2B & G750L & 18-11-27 & 20:00:24 & 1040.0 & odud03040 & 15602 &  \\
G191B2B & G230LB & 21-01-28 & 03:52:02 & 430.0 & oehj03010 & 16436 &  \\
G191B2B & G430L & 21-01-28 & 04:02:47 & 206.0 & oehj03020 & 16436 &  \\
G191B2B & G430L & 21-01-28 & 04:08:23 & 206.0 & oehj03030 & 16436 & E1 \\
G191B2B & G750L & 21-01-28 & 05:22:07 & 968.0 & oehj03040 & 16436 &  \\
GD153 & G230LB & 97-05-21 & 10:18:00 & 600.0 & o3tt42010 & 7063 &  \\
GD153 & G430L & 97-05-21 & 10:34:18 & 252.0 & o3tt42020 & 7063 &  \\
GD153 & G750L & 97-05-21 & 11:31:25 & 3240.0 & o3tt42040 & 7063 &  HCR \\
GD153 & G230LB & 97-05-28 & 06:53:29 & 600.0 & o3tt43010 & 7063 &  \\
GD153 & G430L & 97-05-28 & 07:09:44 & 252.0 & o3tt43020 & 7063 &  \\
GD153 & G750L & 97-05-28 & 08:04:35 & 3240.0 & o3tt43040 & 7063 &  HCR \\
GD153 & G230LB & 97-06-04 & 11:28:46 & 600.0 & o3tt44010 & 7063 &  \\
GD153 & G430L & 97-06-04 & 11:45:01 & 252.0 & o3tt44020 & 7063 &  \\
GD153 & G750L & 97-06-04 & 12:41:08 & 3240.0 & o3tt44040 & 7063 &  HCR \\
GD153 & G230LB & 97-06-10 & 22:22:09 & 600.0 & o3tt45010 & 7063 &  \\
GD153 & G430L & 97-06-10 & 22:38:24 & 252.0 & o3tt45020 & 7063 &  \\
GD153 & G750L & 97-06-10 & 23:33:17 & 3240.0 & o3tt45040 & 7063 &  HCR \\
GD153 & G230LB & 97-06-18 & 04:24:17 & 600.0 & o3tt46010 & 7063 &  \\
GD153 & G430L & 97-06-18 & 04:40:32 & 252.0 & o3tt46020 & 7063 &  \\
GD153 & G750L & 97-06-18 & 05:46:23 & 2282.0 & o3tt46040 & 7063 &  HCR \\
GD153 & G230LB & 97-06-25 & 04:11:40 & 600.0 & o3tt47010 & 7063 &  \\
GD153 & G430L & 97-06-25 & 04:27:55 & 252.0 & o3tt47020 & 7063 &  \\
GD153 & G750L & 97-06-25 & 05:31:49 & 2282.0 & o3tt47040 & 7063 & HCR  \\
GD153 & G230LB & 97-07-01 & 12:03:44 & 600.0 & o3tt48010 & 7063 &  \\
GD153 & G430L & 97-07-01 & 12:19:59 & 252.0 & o3tt48020 & 7063 &  \\
GD153 & G750L & 97-07-01 & 13:11:51 & 2282.0 & o3tt48040 & 7063 &  \\
GD153 & G140L & 97-07-13 & 04:44:14 & 187.1 & o3zx080w0 & 7096 &  \\
GD153 & G230L & 97-07-13 & 04:56:19 & 187.1 & o3zx080v0 & 7096 &  \\
GD153 & G750L & 98-05-17 & 20:42:46 & 900.0 & o4a502020 & 7656 &  G1 \\
GD153 & G430L & 98-05-17 & 21:04:01 & 240.0 & o4a502030 & 7656 &  G1 \\
GD153 & G230LB & 98-05-17 & 22:01:39 & 500.0 & o4a502040 & 7656 &  G1 \\
GD153 & G230L & 98-05-17 & 22:17:27 & 500.0 & o4a502050 & 7656 &  \\
GD153 & G140L & 98-05-17 & 22:33:21 & 500.0 & o4a502060 & 7656 &  \\
GD153 & G230LB & 97-11-12 & 01:28:59 & 400.0 & o4d103010 & 7805 &  \\
GD153 & G430L & 97-11-12 & 01:41:54 & 180.0 & o4d103020 & 7805 &  \\
GD153 & G750L & 97-11-12 & 01:51:09 & 740.0 & o4d103030 & 7805 &  \\
GD153 & G140L & 98-07-15 & 11:07:48 & 204.0 & o4vt100g0 & 8016 &  \\
GD153 & G140L & 98-07-15 & 11:17:26 & 204.0 & o4vt100f0 & 8016 &  P \\
GD153 & G140L & 98-07-15 & 11:28:19 & 204.0 & o4vt100e0 & 8016 &  P \\
GD153 & G140L & 98-07-15 & 11:39:12 & 204.0 & o4vt100d0 & 8016 &  P \\
GD153 & G230L & 98-07-15 & 12:32:53 & 288.0 & o4vt100c0 & 8016 &  \\
GD153 & G230L & 98-07-15 & 12:47:05 & 288.0 & o4vt100b0 & 8016 &  P \\
GD153 & G230L & 98-07-15 & 12:59:22 & 288.0 & o4vt100a0 & 8016 &  P \\
GD153 & G230L & 98-07-15 & 13:11:39 & 288.0 & o4vt10090 & 8016 &  P \\
GD153 & G140L & 02-07-17 & 13:47:04 & 120.0 & o6ig02080 & 8916 &  \\
GD153 & G230L & 02-07-17 & 13:56:29 & 180.0 & o6ig02070 & 8916 &  \\
GD153 & G140L & 02-07-17 & 14:22:13 & 120.0 & o6ig02050 & 8916 &  \\
GD153 & G140L & 01-12-11 & 22:21:36 & 120.0 & o6ig03080 & 8916 &  \\
GD153 & G230L & 01-12-11 & 22:31:01 & 180.0 & o6ig03070 & 8916 &  \\
GD153 & G140L & 01-12-11 & 22:56:45 & 120.0 & o6ig03050 & 8916 &  \\
GD153 & G140L & 02-02-16 & 10:33:09 & 120.0 & o6ig04080 & 8916 &  \\
GD153 & G230L & 02-02-16 & 10:42:34 & 180.0 & o6ig04070 & 8916 &  \\
GD153 & G140L & 02-02-16 & 11:08:18 & 120.0 & o6ig04050 & 8916 &  \\
GD153 & G140L & 02-06-17 & 07:52:12 & 120.0 & o6ig11080 & 8916 &  \\
GD153 & G230L & 02-06-17 & 08:01:37 & 180.0 & o6ig11070 & 8916 &  \\
GD153 & G140L & 02-06-17 & 08:27:21 & 120.0 & o6ig11050 & 8916 &  \\
GD153 & G230LB & 04-01-05 & 08:32:59 & 425.0 & o8v2020g0 & 10039 &  G1 \\
GD153 & G230LB & 04-01-05 & 09:30:44 & 425.0 & o8v2020f0 & 10039 & E1, G1 \\
GD153 & G430L & 04-01-05 & 09:44:11 & 220.0 & o8v2020e0 & 10039 & E1, G1 \\
GD153 & G430L & 04-01-05 & 09:50:01 & 220.0 & o8v2020d0 & 10039 &  G1 \\
GD153 & G750L & 04-01-05 & 10:00:02 & 990.0 & o8v2020c0 & 10039 &  G1 \\
GD153 & G140L & 04-01-05 & 08:01:59 & 450.0 & o8v2020b0 & 10039 &  \\
GD153 & G230L & 04-01-05 & 08:16:54 & 450.0 & o8v2020a0 & 10039 &  \\
GD153 & G140L & 09-08-04 & 22:21:56 & 140.0 & oa8d02010 & 11393 &  \\
GD153 & G140L & 09-08-06 & 16:00:41 & 170.0 & oa9z05010 & 11403 &  \\
GD153 & G230L & 09-08-06 & 16:10:26 & 170.0 & oa9z05020 & 11403 &  \\
GD153 & G230L & 11-01-13 & 23:38:56 & 595.0 & obc402010 & 11999 &  \\
GD153 & G230LB & 11-01-14 & 00:06:09 & 656.0 & obc402030 & 11999 &  \\
GD153 & G430L & 11-01-14 & 01:03:03 & 260.0 & obc402040 & 11999 & E1 \\
GD153 & G430L & 11-01-14 & 01:09:33 & 260.0 & obc402050 & 11999 & HCR \\
GD153 & G750L & 11-01-14 & 01:17:28 & 1980.0 & obc402060 & 11999 &  \\
GD153 & G140L & 11-01-13 & 23:53:29 & 596.0 & obc402enq & 11999 &  \\
GD153 & G230L & 13-01-01 & 15:54:07 & 595.0 & obto10010 & 12682 &  \\
GD153 & G230LB & 13-01-01 & 16:21:18 & 656.0 & obto10030 & 12682 &  \\
GD153 & G430L & 13-01-01 & 17:18:00 & 260.0 & obto10040 & 12682 & E1 \\
GD153 & G430L & 13-01-01 & 17:24:30 & 260.0 & obto10050 & 12682 &  \\
GD153 & G750L & 13-01-01 & 17:32:25 & 1980.0 & obto10060 & 12682 &  \\
GD153 & G140L & 13-01-01 & 16:08:38 & 596.0 & obto10qhq & 12682 &  \\
GD153 & G230L & 13-01-01 & 23:53:01 & 592.0 & oc5506010 & 13162 &  \\
GD153 & G230LB & 13-01-02 & 00:20:09 & 656.0 & oc5506030 & 13162 &  \\
GD153 & G430L & 13-01-02 & 01:16:53 & 260.0 & oc5506040 & 13162 & E1 \\
GD153 & G430L & 13-01-02 & 01:23:23 & 260.0 & oc5506050 & 13162 &  \\
GD153 & G750L & 13-01-02 & 01:31:18 & 1974.0 & oc5506060 & 13162 &  \\
GD153 & G140L & 13-01-02 & 00:07:29 & 596.0 & oc5506rjq & 13162 &  \\
GD153 & G230L & 14-01-20 & 14:31:02 & 592.0 & ocga05010 & 13599 &  \\
GD153 & G230LB & 14-01-20 & 14:58:10 & 656.0 & ocga05030 & 13599 &  \\
GD153 & G430L & 14-01-20 & 15:54:51 & 260.0 & ocga05040 & 13599 & E1 \\
GD153 & G430L & 14-01-20 & 16:01:21 & 260.0 & ocga05050 & 13599 &  \\
GD153 & G750L & 14-01-20 & 16:09:16 & 1974.0 & ocga05060 & 13599 &  \\
GD153 & G140L & 14-01-20 & 14:45:30 & 596.0 & ocga05euq & 13599 &  \\
GD153 & G230L & 17-04-23 & 15:37:47 & 592.0 & odck02010 & 14861 &  \\
GD153 & G230LB & 17-04-23 & 16:04:51 & 638.0 & odck02030 & 14861 &  \\
GD153 & G430L & 17-04-23 & 17:03:08 & 254.0 & odck02040 & 14861 & E1 \\
GD153 & G430L & 17-04-23 & 17:09:32 & 256.0 & odck02050 & 14861 &  \\
GD153 & G750L & 17-04-23 & 17:17:23 & 1962.0 & odck02060 & 14861 &  \\
GD153 & G140L & 17-04-23 & 15:52:15 & 592.0 & odck02p2q & 14861 &  \\
GD153 & G230L & 19-05-13 & 23:17:50 & 580.0 & odud02010 & 15602 &  \\
GD153 & G230LB & 19-05-13 & 23:44:31 & 596.0 & odud02030 & 15602 &  \\
GD153 & G430L & 19-05-14 & 00:44:29 & 254.0 & odud02040 & 15602 & E1 \\
GD153 & G430L & 19-05-14 & 00:50:53 & 256.0 & odud02050 & 15602 &  \\
GD153 & G750L & 19-05-14 & 00:58:44 & 1818.0 & odud02060 & 15602 &  \\
GD153 & G140L & 19-05-13 & 23:32:06 & 581.0 & odud02fpq & 15602 &  \\
GD153 & G230L & 21-01-25 & 14:29:36 & 565.0 & oehj02010 & 16436 &  \\
GD153 & G230LB & 21-01-25 & 14:55:47 & 566.0 & oehj02030 & 16436 &  \\
GD153 & G430L & 21-01-25 & 15:56:48 & 250.0 & oehj02040 & 16436 & E1 \\
GD153 & G430L & 21-01-25 & 16:03:08 & 250.0 & oehj02050 & 16436 &  \\
GD153 & G750L & 21-01-25 & 16:10:53 & 1737.0 & oehj02060 & 16436 &  \\
GD153 & G140L & 21-01-25 & 14:43:37 & 566.0 & oehj02gyq & 16436 &  \\
GD71 & G750L & 98-02-14 & 13:10:14 & 1980.0 & o49x09010 & 7674 &  \\
GD71 & G750L & 98-03-17 & 12:59:21 & 1980.0 & o49x10010 & 7674 &  \\
GD71 & G750L & 99-04-23 & 17:08:53 & 900.0 & o4a520020 & 7656 &  G1 \\
GD71 & G430L & 99-04-23 & 17:30:14 & 240.0 & o4a520030 & 7656 &  G1 \\
GD71 & G230LB & 99-04-23 & 18:27:22 & 500.0 & o4a520040 & 7656 &  G1 \\
GD71 & G230L & 99-04-23 & 18:43:05 & 500.0 & o4a520050 & 7656 &  \\
GD71 & G140L & 99-04-23 & 18:59:14 & 500.0 & o4a520060 & 7656 &  \\
GD71 & G750L & 98-11-04 & 08:10:14 & 900.0 & o4a551020 & 7656 &  G1 \\
GD71 & G430L & 98-11-04 & 08:31:29 & 240.0 & o4a551030 & 7656 &  G1 \\
GD71 & G230LB & 98-11-04 & 09:30:28 & 500.0 & o4a551040 & 7656 &  G1 \\
GD71 & G230L & 98-11-04 & 09:46:11 & 500.0 & o4a551050 & 7656 &  \\
GD71 & G140L & 98-11-04 & 10:02:15 & 500.0 & o4a551060 & 7656 &  \\
GD71 & G140L & 98-03-31 & 15:58:37 & 96.0 & o4pg010l0 & 7917 &  \\
GD71 & G140L & 98-03-31 & 16:07:32 & 108.0 & o4pg010k0 & 7917 & \\
GD71 & G140L & 98-03-31 & 16:16:39 & 108.0 & o4pg010j0 & 7917 & P \\
GD71 & G140L & 98-03-31 & 17:06:45 & 108.0 & o4pg010i0 & 7917 &  P \\
GD71 & G140L & 98-03-31 & 17:15:52 & 108.0 & o4pg010h0 & 7917 &  P \\
GD71 & G230L & 98-03-31 & 17:27:20 & 216.0 & o4pg010g0 & 7917 &  \\
GD71 & G140L & 98-04-27 & 06:15:01 & 108.0 & o4sp01060 & 7932 &   \\
GD71 & G140L & 99-02-06 & 14:43:39 & 72.0 & o53001080 & 7937 &  P \\
GD71 & G140L & 00-02-29 & 12:55:30 & 500.0 & o5i001010 & 8421 &  P \\
GD71 & G230L & 00-02-29 & 13:10:45 & 500.0 & o5i001020 & 8421 &  P \\
GD71 & G230LB & 00-02-29 & 13:24:09 & 500.0 & o5i001030 & 8421 &  G1 \\
GD71 & G430L & 00-02-29 & 14:20:21 & 240.0 & o5i001040 & 8421 &  G1 \\
GD71 & G750L & 00-02-29 & 14:30:42 & 900.0 & o5i001050 & 8421 &  G1 \\
GD71 & G140L & 00-03-01 & 06:40:22 & 108.0 & o5i0020l0 & 8421 &  \\
GD71 & G140L & 00-03-01 & 06:48:04 & 108.0 & o5i0020k0 & 8421 &  P \\
GD71 & G140L & 00-03-01 & 06:57:01 & 108.0 & o5i0020j0 & 8421 &  P \\
GD71 & G140L & 00-03-01 & 07:05:58 & 108.0 & o5i0020i0 & 8421 &   \\
GD71 & G140L & 00-03-01 & 07:14:55 & 108.0 & o5i0020h0 & 8421 &  P \\
GD71 & G230L & 00-03-01 & 08:04:50 & 216.0 & o5i0020g0 & 8421 &  \\
GD71 & G430L & 00-01-15 & 07:35:38 & 240.0 & o61001010 & 8505 &  G1 \\
GD71 & G430L & 00-01-15 & 07:41:06 & 240.0 & o61001020 & 8505 &  \\
GD71 & G230LB & 00-01-15 & 07:51:27 & 700.0 & o61001030 & 8505 &  G1 \\
GD71 & G750L & 00-01-15 & 08:10:28 & 700.0 & o61001040 & 8505 &  G1 \\
GD71 & G140L & 00-01-15 & 09:19:23 & 250.0 & o61001070 & 8505 &  \\
GD71 & G230L & 00-01-15 & 09:30:28 & 240.0 & o61001080 & 8505 &  \\
GD71 & G430L & 00-01-18 & 06:39:11 & 240.0 & o61002010 & 8505 &  G1 \\
GD71 & G230LB & 00-01-18 & 06:49:32 & 700.0 & o61002020 & 8505 &  G1 \\
GD71 & G750L & 00-01-18 & 07:07:33 & 700.0 & o61002030 & 8505 &  G1, HCR \\
GD71 & G140L & 00-01-18 & 08:27:02 & 250.0 & o61002060 & 8505 &  P \\
GD71 & G230L & 00-01-18 & 08:38:07 & 250.0 & o61002070 & 8505 &  \\
GD71 & G430L & 00-01-20 & 10:11:22 & 240.0 & o61003010 & 8505 &  G1 \\
GD71 & G230LB & 00-01-20 & 10:21:43 & 700.0 & o61003020 & 8505 &  G1 \\
GD71 & G750L & 00-01-20 & 10:39:44 & 700.0 & o61003030 & 8505 &  G1, HCR \\
GD71 & G140L & 00-01-20 & 11:57:10 & 250.0 & o61003060 & 8505 &  \\
GD71 & G230L & 00-01-20 & 12:08:15 & 250.0 & o61003070 & 8505 &  \\
GD71 & G430L & 00-01-27 & 06:25:30 & 240.0 & o61004010 & 8505 & G1 \\
GD71 & G230LB & 00-01-27 & 06:35:51 & 700.0 & o61004020 & 8505 &  G1 \\
GD71 & G750L & 00-01-27 & 06:53:52 & 700.0 & o61004030 & 8505 &  G1, HCR \\
GD71 & G140L & 00-01-27 & 08:10:04 & 250.0 & o61004060 & 8505 &  \\
GD71 & G230L & 00-01-27 & 08:21:09 & 250.0 & o61004070 & 8505 &  \\
GD71 & G430L & 00-02-01 & 21:25:52 & 240.0 & o61005010 & 8505 & G1 \\
GD71 & G230LB & 00-02-01 & 21:36:13 & 700.0 & o61005020 & 8505 & G1 \\
GD71 & G750L & 00-02-01 & 21:54:14 & 700.0 & o61005030 & 8505 & G1 \\
GD71 & G140L & 00-02-01 & 23:06:18 & 250.0 & o61005060 & 8505 &  \\
GD71 & G230L & 00-02-01 & 23:17:23 & 250.0 & o61005070 & 8505 &  \\
GD71 & G430L & 00-02-06 & 20:48:44 & 240.0 & o61006010 & 8505 & G1 \\
GD71 & G230LB & 00-02-06 & 20:59:05 & 700.0 & o61006020 & 8505 & G1 \\
GD71 & G750L & 00-02-06 & 21:17:06 & 700.0 & o61006030 & 8505 &  G1, HCR \\
GD71 & G140L & 00-02-06 & 22:33:09 & 250.0 & o61006060 & 8505 & \\
GD71 & G230L & 00-02-06 & 22:44:14 & 250.0 & o61006070 & 8505 &  \\
GD71 & G230LB & 01-12-07 & 05:41:27 & 310.0 & o6ig010i0 & 8916 &  G1 \\
GD71 & G430L & 01-12-07 & 05:52:58 & 150.0 & o6ig010h0 & 8916 &  G1 \\
GD71 & G750L & 01-12-07 & 06:01:49 & 440.0 & o6ig010g0 & 8916 &  G1 \\
GD71 & G140L & 01-12-07 & 02:48:39 & 450.0 & o6ig010f0 & 8916 &  \\
GD71 & G230L & 01-12-07 & 03:03:34 & 500.0 & o6ig010e0 & 8916 &  \\
GD71 & G430L & 03-10-24 & 01:29:17 & 180.0 & o8v2010k0 & 10039 &  G1 \\
GD71 & G430L & 03-10-24 & 01:34:27 & 180.0 & o8v2010j0 & 10039 & E1, G1 \\
GD71 & G750L & 03-10-24 & 02:59:17 & 720.0 & o8v2010i0 & 10039 &  G1 \\
GD71 & G140L & 03-10-24 & 00:59:20 & 380.0 & o8v2010h0 & 10039 &  \\
GD71 & G230L & 03-10-24 & 02:29:05 & 360.0 & o8v2010e0 & 10039 &  \\
GD71 & G230LB & 03-10-24 & 04:42:17 & 350.0 & o8v204040 & 10039 & G1 \\
GD71 & G230LB & 03-10-24 & 07:14:04 & 350.0 & o8v204080 & 10039 & E1, G1 \\
GD71 & G230L & 10-03-10 & 03:38:31 & 595.0 & obc401010 & 11999 &  \\
GD71 & G230LB & 10-03-10 & 04:05:48 & 660.0 & obc401030 & 11999 &  \\
GD71 & G430L & 10-03-10 & 05:00:47 & 260.0 & obc401040 & 11999 & E1 \\
GD71 & G430L & 10-03-10 & 05:07:16 & 260.0 & obc401050 & 11999 &  \\
GD71 & G750L & 10-03-10 & 05:15:11 & 2010.0 & obc401060 & 11999 &  \\
GD71 & G140L & 10-03-10 & 03:53:05 & 600.0 & obc401m5q & 11999 &  \\
GD71 & G230L & 11-11-02 & 21:01:37 & 595.0 & obvp06010 & 12737 &  \\
GD71 & G230LB & 11-11-02 & 21:29:05 & 642.0 & obvp06030 & 12737 & E1 \\
GD71 & G430L & 11-11-02 & 22:26:34 & 260.0 & obvp06040 & 12737 & E1 \\
GD71 & G430L & 11-11-02 & 22:33:04 & 260.0 & obvp06050 & 12737 &  \\
GD71 & G750L & 11-11-02 & 22:40:59 & 1980.0 & obvp06060 & 12737 & HCR \\
GD71 & G140L & 11-11-02 & 21:16:10 & 595.0 & obvp06a8q & 12737 &  \\
GD71 & G230LB & 13-02-12 & 12:32:00 & 600.0 & oc3i15010 & 12813 & E1 \\
GD71 & G430L & 13-02-12 & 12:45:35 & 308.0 & oc3i15020 & 12813 & E1 \\
GD71 & G750L & 13-02-12 & 12:55:00 & 1000.0 & oc3i15030 & 12813 &  \\
GD71 & G230L & 14-02-05 & 15:52:07 & 595.0 & ocga04010 & 13599 &  \\
GD71 & G230LB & 14-02-05 & 16:19:33 & 640.0 & ocga04030 & 13599 & E1 \\
GD71 & G430L & 14-02-05 & 17:14:06 & 260.0 & ocga04040 & 13599 & E1 \\
GD71 & G430L & 14-02-05 & 17:20:37 & 260.0 & ocga04050 & 13599 &  \\
GD71 & G750L & 14-02-05 & 17:28:31 & 1974.0 & ocga04060 & 13599 & HCR \\
GD71 & G140L & 14-02-05 & 16:06:37 & 595.0 & ocga04twq & 13599 &  \\
GD71 & G230L & 17-01-17 & 22:15:52 & 595.0 & odck01010 & 14861 &  \\
GD71 & G230LB & 17-01-17 & 22:43:18 & 616.0 & odck01030 & 14861 & E1 \\
GD71 & G430L & 17-01-17 & 23:41:16 & 254.0 & odck01040 & 14861 & E1 \\
GD71 & G430L & 17-01-17 & 23:47:40 & 256.0 & odck01050 & 14861 &  \\
GD71 & G750L & 17-01-17 & 23:55:31 & 1962.0 & odck01060 & 14861 & HCR  \\
GD71 & G140L & 17-01-17 & 22:30:23 & 595.0 & odck01mwq & 14861 &  \\
GD71 & G230L & 19-03-28 & 17:42:04 & 582.0 & odud01010 & 15602 &  \\
GD71 & G230LB & 19-03-28 & 18:09:02 & 580.0 & odud01030 & 15602 & E1 \\
GD71 & G430L & 19-03-28 & 19:08:46 & 254.0 & odud01040 & 15602 & E1 \\
GD71 & G430L & 19-03-28 & 19:15:10 & 254.0 & odud01050 & 15602 &  \\
GD71 & G750L & 19-03-28 & 19:22:59 & 1818.0 & odud01060 & 15602 & HCR \\
GD71 & G140L & 19-03-28 & 17:56:22 & 580.0 & odud01r1q & 15602 &  \\
GD71 & G230L & 21-03-16 & 21:28:52 & 560.0 & oehj01010 & 16436 &  \\
GD71 & G230LB & 21-03-16 & 21:55:08 & 562.0 & oehj01030 & 16436 & E1 \\
GD71 & G430L & 21-03-16 & 22:56:01 & 250.0 & oehj01040 & 16436 & E1 \\
GD71 & G430L & 21-03-16 & 23:02:21 & 250.0 & oehj01050 & 16436 &  \\
GD71 & G750L & 21-03-16 & 23:10:06 & 1737.0 & oehj01060 & 16436 & HCR  \\
GD71 & G140L & 21-03-16 & 21:42:48 & 560.0 & oehj01izq & 16436 &  \\    
    \enddata
\end{deluxetable}

\end{document}